\documentclass{IEEEtaes}
\IEEEoverridecommandlockouts
\IEEEoverridecommandlockouts
\usepackage{cite}
\usepackage{amsmath,amssymb,amsfonts}
\usepackage{mathrsfs}
\usepackage{algorithmic}
\usepackage[linesnumbered,ruled]{algorithm2e}
\usepackage{multirow}
\usepackage{booktabs}
\usepackage{soul}
\usepackage[normalem]{ulem}

\usepackage{psfrag}
\usepackage[table]{xcolor}
\usepackage{color}
\usepackage{graphicx}
\usepackage{subfig}
\usepackage{float}
\usepackage{caption}
\usepackage{subcaption}
\usepackage{acronym}
\usepackage{xcolor}
\usepackage{comment}
\usepackage{longtable}
\usepackage{booktabs, tabularx}
\usepackage{tikz}
\usepackage{pgfplots}
\usepackage{bbm}

\definecolor{darkred}{RGB}{128, 0, 34}

\SetKwFor{For}{for}{do}{end for}
\SetKwIF{If}{ElseIf}{Else}{if}{then}{else if}{else}{end if}
\SetAlgoLined

\begin{document}

\title{Human-Inspired Pavlovian and Instrumental Learning for Autonomous Agents Navigation}

\author{Jingfeng Shan}
\member{Member, IEEE}
\affil{University of Bologna, Bologna, Italy} 

\author{Francesco Guidi}
\member{Member, IEEE}
\affil{National Research Council of Italy, Bologna, Italy}

\author{Mehrdad Saeidi}
\member{Member, IEEE}
\affil{University of Bologna, Bologna, Italy} 

\author{Enrico Testi}
\member{Member, IEEE}
\affil{University of Bologna, Bologna, Italy}

\author{Elia Favarelli}
\member{Member, IEEE}
\affil{University of Bologna, Bologna, Italy}

\author{Andrea Giorgetti}
\member{Senior Member, IEEE}
\affil{University of Bologna, Cesena, Italy}

\author{Davide Dardari}
\member{Fellow, IEEE}
\affil{University of Bologna, Cesena, Italy}

\author{Alberto Zanella}
\member{Senior Member, IEEE}
\affil{National Research Council of Italy, Bologna, Italy}

\author{Giorgio Li Pira}
\affil{ University of Bologna, Bologna, Italy}

\author{Francesca Starita}
\affil{ University of Bologna, Bologna,  Italy}

\author{Anna Guerra}
\member{Member, IEEE}
\affil{University of Bologna, Cesena, Italy}

\receiveddate{This work was partially supported by the European Union under the NRPP of  NextGenerationEU (Mision 4  – Component 2  -Investment 1.1) Prin 2022 (No. 104, 2/2/2022, CUP J53C24002790006), under ERC Grant no. 101116257 (project CUE-GO: Contextual Radio Cues for Enhancing Decision Making in Networks of Autonomous Agents).
}

\corresp{{\itshape (Corresponding author: J, Shan)}. }

\authoraddress{J. Shan, M. Saeidi, E. Testi, E. Favarelli, D. Dardari, and A. Guerra are with WiLab, University of Bologna, via dell'Universit\'a 50, 47521 Cesena, Italy, e-mail: \{enrico.testi, elia.favarelli, andrea.giorgetti, davide.dardari, anna.guerra3\}@unibo.it.
F. Guidi and A. Zanella are with the National Research Council of Italy (CNR), viale del Risorgimento 2, 40136 Bologna, Italy, e-mail: \{francesco.guidi, alberto.zanella\}@cnr.it.
G. Li Pira and F. Starita are with the University of Bologna, Department of Psychology ``Renzo Canestrari", 47521 Cesena, Italy.}

% The paper headers

\markboth{Submitted to IEEE Transactions on Aerospace and Electronic Systems}%
{J. Shan  \MakeLowercase{\textit{et al.}}}

\maketitle

%\newpage 
\begin{abstract}
Autonomous agents operating in uncertain environments must balance fast responses with goal-directed planning. Classical \ac{MF} \ac{RL} often converges slowly and may induce unsafe exploration, whereas \ac{MB} methods are computationally expensive and sensitive to model mismatch. This paper presents a human-inspired hybrid \ac{RL} architecture integrating Pavlovian, Instrumental \ac{MF}, and Instrumental \ac{MB} components. Inspired by Pavlovian and Instrumental learning from neuroscience, the framework considers contextual radio cue, here intended as georeferenced environmental features acting as \ac{CS}, to shape intrinsic value signals and bias decision-making. Learning is further modulated by internal motivational drives through a dedicated motivational signal. A Bayesian arbitration mechanism adaptively blends \ac{MF} and \ac{MB} estimates based on predicted reliability.
Simulation results show that the hybrid approach accelerates learning, improves operational safety, and reduces navigation in high-uncertainty regions compared to standard RL baselines. Pavlovian conditioning promotes safer exploration and faster convergence, while arbitration enables a smooth transition from exploration to efficient, plan-driven exploitation. Overall, the results highlight the benefits of biologically inspired modularity for robust and adaptive autonomous systems under uncertainty.
\end{abstract}


\acresetall
\bstctlcite{IEEEexample:BSTcontrol}

% new commands - Jingfeng
\newcommand{\Vp} {V^{\mathrm{P},i}}
\newcommand{\Qi} {Q^{\mathrm{I},i}}
\newcommand{\Vi} {V^{\mathrm{I},i}}
\newcommand{\rpt} {r^{\mathrm{P}}_{i,t}}
\newcommand{\rit} {r^{\mathrm{I}}_{i,t}}
\newcommand{\gammaV} {\gamma_\mathrm{V} }
\newcommand{\gammaQ} {\gamma_\mathrm{Q} }
\newcommand{\alphaV} {\alpha_\mathrm{V} }
\newcommand{\alphaQ} {\alpha_\mathrm{Q} }
\newcommand{\wo} {w_1}
\newcommand{\wt} {w_2}
\newcommand{\att} {a_{i,t}}
\newcommand{\atto} {a_{i,t+1}}
\newcommand{\stt} {s_{i,t}}
\newcommand{\stto} {s_{i,t+1}}
\newcommand{\Mit} {M_{i,t}}
\newcommand{\bit} {\bar{b}_{i,t}}
\newcommand{\tauit} {\bar{\tau}_{i,t}}

% A
\acrodef{AA}{autonomous agent}
\acrodef{AI}{artificial intelligence}
\acrodef{AOA}{angle-of-arrival}
\acrodef{A2A}{agent-to-agent}
% C
\acrodef{COA}{curvature-of-arrival}
\acrodef{CS}{conditioned stimulus}
\acrodef{CR}{conditioned response}
\acrodef{CSI}{channel state information}
\acrodef{CRLB}{Cram\'er-Rao Lower Bound}
\acrodef{CDF}{cumulative density function}
% D
\acrodef{DRN}{dynamic radar network}
%E
\acrodef{EKF}{extended Kalman filter}
\acrodef{EM}{electromagnetic}
\acrodef{EA}{evolutionary algorithm}
\acrodef{ECDF}{empirical cumulative density function}
%F
\acrodef{FIM}{Fisher Information Matrix}
%G
\acrodef{GA}{genetic algorithm}
\acrodef{GT}{goal-tracker}
%K
\acrodef{KF}{Kalman filter}
\acrodef{KL}{Kullback-Leibler}
\acrodef{KS}{Kolmogorov-Smirnov}
% L
\acrodef{LS}{least-squares}
\acrodef{LOS}{line-of-sight}
%M
\acrodef{MPB}{measurement prediction bound}
\acrodef{MDP}{Markov decision process}
\acrodef{MLE}{maximum likelihood estimator}
\acrodef{MAP}{maximum {\em a posteriori}}
\acrodef{MMSE}{minimum mean square error}
\acrodef{MSE}{mean square error}
\acrodef{MF}{Model-Free}
\acrodef{MB}{Model-Based}
%N
\acrodef{NLOS}{non-line-of-sight}
\acrodef{NS}{neutral stimulus}
%P
\acrodef{PF}{particle filter}
\acrodef{PSGD}{projected steepest gradient descent}
\acrodef{PDF}{probability density function}
\acrodef{PSO}{particle swarm optimization}
\acrodef{PSA}{proficiency self-assessment}
\acrodef{PIT}{Pavlovian-Instrumental Transfer}
\acrodef{PEB}{Position Error Bound}
% R
\acrodef{RSSI}{received signal strength indicator}
\acrodef{RMSE}{root mean square error}
\acrodef{RPE}{reward prediction error}
\acrodef{RV}{random variable}
\acrodef{RL}{reinforcement learning}
%S
\acrodef{SAR}{synthetic aperture radar}
\acrodef{SPE}{state prediction error}
\acrodef{SSM}{State-Space Model}
\acrodef{SLAM}{simultaneous localization and mapping}
\acrodef{SNR}{signal-to-noise ratio}
\acrodef{ST}{sign-tracker}
%T
\acrodef{TD}{temporal-difference}
\acrodef{ToA}{time-of-arrival}
%U
\acrodef{UAV}{unmanned aerial vehicle}
\acrodef{UKF}{unscented Kalman filter}
\acrodef{UPF}{unscented particle filter}
\acrodef{US}{unconditioned stimulus}

%\tableofcontents
\section{Introduction}
Autonomous agents such as \acp{UAV}, mobile robots, and autonomous ground vehicles are increasingly deployed in mission-critical applications, including post-disaster scenarios \cite{guerra2023reinforcement}. These environments are characterized by high uncertainty, degraded communication infrastructures, and stringent real-time constraints \cite{GuvEtAl:J18,GueEtAl:J22, Azzam2024}. Agents must operate on noisy, high-dimensional sensory data, adapt to unpredictable disturbances, and make decisions that balance immediate task demands with long-term objectives. A core challenge in autonomous agent networks is transforming uncertain observations into reliable state estimates and mapping them to control actions that satisfy safety, efficiency, and performance requirements under limited computational and energy resources \cite{DorTheTri:J21, Alexandropoulos2025, Soleymani2025}.

Most autonomous decision-making approaches rely on \ac{MF} \ac{RL}, where agents iteratively refine policies through reward feedback and environmental interaction \cite{9362239, lee2020optimization, lahmeri2021artificial,fontanesi2025deep}. While effective in controlled settings, these methods treat learning as a monolithic optimization process driven by uniform \ac{TD} updates. This perspective contrasts with biological decision-making systems, where multiple functionally distinct learning mechanisms coexist and interact \cite{6316049, dayan2002reward, hassabis2017neuroscience}. Such parallel systems contribute differently to behavioral control, enabling faster learning while balancing safety and efficiency in unfamiliar and potentially hazardous environments.

Neuroscientific evidence indicates that human decision-making is governed by at least two dissociable learning systems \cite{afshar2023reward, ghirlanda2020learning, tsurumi2025online}. \emph{Instrumental conditioning} supports goal-directed behavior by learning action–outcome contingencies, reinforcing actions that lead to favorable outcomes while suppressing harmful ones \cite{dayan2014model}. In contrast, \emph{Pavlovian conditioning} learns stimulus–outcome associations, allowing organisms to anticipate rewards or punishments based on predictive environmental cues, independently of their actions \cite{delamater2007learning, robinson2013instant, pool2019behavioural}. Pavlovian learning enables fast, reflexive responses to contextual cues, which in engineered systems may include radio-derived environmental features, supporting anticipatory behavior in safety-critical scenarios. Importantly, Pavlovian and instrumental systems operate in parallel, with Pavlovian predictions biasing instrumental action selection toward expected outcomes.

Motivated by these insights, the authors in \cite{lee2019decision} advocated closer integration between neuroscience and robotics, proposing a conceptual hierarchical architecture comprising reflexes, Pavlovian responses, \ac{MF} habitual control, and \ac{MB} goal-directed planning, coordinated by an arbitration mechanism sensitive to uncertainty. However, their contribution remains largely at a theoretical level, without specifying a computational implementation or a concrete learning algorithm for autonomous agents.

Then, \cite{mahajan2024balancing} proposed the PAL algorithm that integrates Pavlovian fear signals into standard \ac{RL} through uncertainty, modulated action biases. The MaxPain framework \cite{elfwing2017parallel} separates reward and punishment learning via distinct $Q$-functions, yielding more conservative exploration, while its extension in \cite{wang2021modular} introduces signal-specific discounting and adaptive weighting of subsystems.

Despite these advances, it is still unclear how Pavlovian stimulus–outcome predictions interact with instrumental action–outcome learning to shape decision-making. This interaction is named in neuroscience as \ac{PIT} \cite{cartoni2013three, cartoni2013bayesian}, whereby Pavlovian conditioned stimuli modulate instrumental behavior. \ac{PIT} manifests in specific, general, and inhibitory forms, influencing action selection even when cues provide no information about the optimal action. This highlights a deep interaction between prediction-based and control-based learning mechanisms rather than simple information fusion.

From a computational perspective, \ac{PIT} poses a significant challenge to standard actor–critic architectures, which conflate state-value estimation and policy optimization and therefore fail to capture several key \ac{PIT} phenomena \cite{dayan2002reward}. In particular, such models struggle to account for reflexive, stimulus-driven responses that occur independently of action–outcome contingencies, for the persistent modulation of behavior by internal motivational states even after extensive training, and for Pavlovian biases that can systematically influence action selection in ways that may be suboptimal from a purely instrumental standpoint.

The advantage function, defined as the difference between state-action values and state values, has been proposed as a natural computational substrate for modeling \ac{PIT} effects in computational neuroscience \cite{dayan2002reward, 374604, dayan1992q}. Recent works have leveraged advantage-based formulations in multi-agent and robotic systems \cite{jung2025multi, li2024continuous} to improve convergence and coordination. Within this framework, the state-value component can be interpreted as a Pavlovian predictor, while action-specific advantages encode instrumental control. As learning progresses and optimal actions are repeatedly selected, advantages diminish, yielding a transition from deliberative control to automatic, value-driven responding, thus mirroring the shift from goal-directed to habitual behavior under Pavlovian influence \cite{dayan2002reward}.

Unfortunately, advantage-based \ac{PIT} models remain largely unexplored in the design of autonomous navigation of agents. As an example, in the context of target localization under GPS-denied navigation, agents must rapidly exploit contextual environmental cues, including radio signal features, to anticipate risks and opportunities. Learning predictive associations between such cues and task-relevant outcomes (e.g., localization accuracy) could substantially enhance robustness and adaptivity. Moreover, explicitly separating Pavlovian and instrumental processes may enable heterogeneous multi-agent systems in which agents exhibit complementary behaviors, improving collective performance in complex environments.

Building on these insights, and on \cite{guidi2026cognitive} where we provide a first discussion on the human-inspired architecture, this work introduces a novel Pavlovian–Instrumental \ac{RL} architecture, in which \ac{PIT} mechanisms are explicitly realized within a \ac{MF} learning framework, while a complementary hybrid \ac{MB}/\ac{MF} component provides goal-directed planning through arbitration, thereby enhancing autonomous decision-making in unknown environments.

The main contributions of this paper are summarized as follows.
\begin{itemize}
\item We propose a human-inspired decision-making framework integrating three learning systems: a Pavlovian module for cue-based modulation, an Instrumental \ac{MF} learner for habitual control, and an Instrumental \ac{MB} planner for goal-directed planning.
\item We introduce the concept of \emph{contextual environmental cues}, defined as georeferenced features that act as \ac{CS} and generate intrinsic Pavlovian value signals, thereby biasing action selection toward information-rich regions and away from hazardous areas. The proposed framework is general and can accommodate different types of environmental cues. In this work, we focus on radio-based cues as a representative example of practical interest for wireless localization, and adopt them in the numerical results.

\item We propose a novel learning architecture that integrates Pavlovian and instrumental components. Pavlovian predictions associated with environmental cues are converted, through a dedicated modulation function, into action-dependent approach-avoidance biases that influence the agent's action selection. A motivational gate further modulates learning dynamics based on internal state variables, such as battery level and mission time.
\item We validate the proposed framework in a multi-agent target localization task, demonstrating improved performance compared to standard \ac{RL} baselines.
\end{itemize}

The remainder of this paper is organized as follows. Section \ref{sec:architecture} defines the problem formulation, reviews the biological background of instrumental and Pavlovian conditioning, and presents the proposed architecture. Sections \ref{sec:methodology} describes the considered framework, while Sections~\ref{sec: hybrid}  reports the proposed algorithm and the hybrid approach. Section~\ref{sec: casestudy_vector} describes simulation results, while final conclusions are drawn in Section~\ref{sec:conclusions}. Table \ref{tab:neuro_acronyms} summarizes the main acronyms from behavioral neuroscience used throughout the Pavlovian framework.

\section{Human-Inspired Digital Agent Architecture}
\label{sec:architecture}

\subsection{Research Problem}
\label{sec:problem}
Autonomous agents deployed in complex environments must meet several, often
conflicting, requirements: efficient exploration, safe navigation, rapid adaptation to
uncertain sensory conditions, and effective long-horizon reasoning. Traditional
\ac{MF} \ac{RL} methods often struggle in such settings,
as they rely solely on reward-driven trial-and-error and lack mechanisms for
leveraging predictive environmental cues or expressing innate behavioral
biases.

Neuroscientific evidence demonstrates that biological decision-making is governed by the interaction of multiple dissociable learning systems, including Pavlovian, \ac{MF}, and \ac{MB} controllers, which contribute unique strengths spanning rapid reactivity, robustness to uncertainty, and long-term foresight \cite{dayan2002reward,lee2019decision}. Yet, despite substantial work on biologically inspired RL \cite{6316049}, a principled integration of these subsystems remains underexplored in autonomous agents,
especially in scenarios where contextual cues (e.g., radio features) carry task-related meaning.

This paper addresses the following research question:

\emph{How can Pavlovian, instrumental \ac{MF}, and \ac{MB} learning be integrated into a unified decision architecture that enables autonomous agents to navigate and localize targets efficiently in unknown environments?}
\begin{table}[!t]
\caption{Neuroscience Acronyms}
\label{tab:neuro_acronyms}
\centering
\begin{tabular}{ll}
\hline
Acronym & Meaning \\
\hline
NS & Neutral Stimulus \\
CS & Conditioned Stimulus \\
US & Unconditioned Stimulus \\
UR & Unconditioned Response \\
CR & Conditioned Response \\
PIT & Pavlovian-to-Instrumental Transfer \\
\hline
\end{tabular}
\end{table}

To tackle this problem, we develop a human-inspired hybrid \ac{RL} algorithm that: (i) incorporates Pavlovian value signals derived from contextual radio cues, (ii) maintains instrumental value estimates through both \ac{MF} and \ac{MB}
updates, and (iii) dynamically arbitrates between \ac{MF} and \ac{MB} control using a Bayesian reliability mechanism. 

The proposed framework is evaluated in a
multi-agent  localization task featuring gates, obstacles, and GPS-denied
regions, where anticipatory cue-guided behavior is essential for safe and
efficient mission performance.

\subsection{Background on Instrumental and Pavlovian Conditioning}

Pavlovian and instrumental conditioning constitute two distinct learning mechanisms with complementary computational roles \cite{sutton1998reinforcement, dayan2002reward}. Pavlovian conditioning provides fast, reflexive responses driven by stimulus–outcome associations, whereas instrumental conditioning supports flexible action–outcome learning and policy optimization. Their interaction is central to designing agents that combine rapid, cue-driven behavior with goal-directed control.

\paragraph*{Pavlovian Conditioning}
In animals, Pavlovian conditioning involves learning predictive relationships between environmental cues and biologically significant outcomes. Through repeated pairings between a \ac{NS} and an \ac{US}, {(i.e. a biologically relevant stimulus such as food or pain),} the \ac{NS} becomes a \ac{CS} capable of triggering a \ac{CR}. A classic example is Pavlov’s experiment, where a neutral auditory cue paired with food became sufficient to elicit salivation.

Pavlovian responses are \emph{automatic} and rely on evolutionarily specified stimulus–response mappings rather than action selection. According to \cite{dayan2002reward}, these responses reflect both (i) a hard-wired motivational system, immediately sensitive to internal states (e.g., hunger, thirst), and (ii) a learned component mediated by \ac{CS}–\ac{US} associations.

\paragraph*{Instrumental Conditioning}
Instrumental conditioning concerns learning action–outcome relationships. It relies on two systems: (i) \emph{Habitual (\ac{MF}):} acquires stimulus–response associations via reinforcement and {often becomes insensitive to outcome revaluation following extensive experience with the association (so called, overtraining)}; (ii) \emph{Goal-directed (\ac{MB}):} selects actions by evaluating predicted consequences and {is influenced by the agent’s internal motivational state, ensuring that instrumental actions are inhibited when the predicted outcome no longer carries incentive value (e.g., the absence of hunger suppresses the pursuit of food) \cite{dayan2002reward}.

\paragraph*{Pavlovian–Instrumental Transfer (PIT)}
Pavlovian cues can energize or bias instrumental actions, a phenomenon known as \ac{PIT} \cite{cartoni2013three}. The authors in \cite{dayan2002reward} highlighted that \ac{PIT} is driven by the Pavlovian motivational system and is neurally dissociable from goal-directed incentive learning. Consequently, Pavlovian value can modulate instrumental policies even in the absence of direct action–outcome information. This mechanism motivates the hybrid control scheme adopted in this work, where Pavlovian value modulates advantage-based instrumental decision making, enabling both rapid reactivity and long-horizon planning.

\subsection{Human-Inspired Learning Architecture}

Fig.~\ref{fig:framework} illustrates the proposed human-inspired hybrid learning architecture, which governs how an autonomous agent interacts with its environment and internally regulates perception, learning, and action selection. The design is conceptually inspired by biological decision-making frameworks such as \cite{lee2019decision}, in which multiple learning systems operate in parallel and are dynamically combined based on their reliability and contextual demands. 

Unlike the biological framework in \cite{lee2019decision}, which focuses on modeling neural decision mechanisms, the proposed architecture translates these principles into a computational control framework for autonomous digital agents. In particular, it introduces (i) contextual environmental cues derived from radio features, (ii) a Pavlovian mechanism that generates action-dependent approach–avoidance biases, and (iii) a motivational gating mechanism that modulates learning according to operational constraints such as battery level and mission time.

\subsubsection{External Perception--Action Loop}

The perception--action loop constitutes the interaction between the agent and the external environment. At each timestep, the agent acquires sensory inputs from onboard sensors (e.g., radio measurements, cameras), which encode partial and noisy information about the environment. These observations are processed through learning and decision processes.
Actions generated by the decision module alter the environment and yield new sensory inputs or external reinforcement signals, thereby closing the loop. This mechanism corresponds to the top-level flow in Fig.~\ref{fig:framework} (dashed lines), where perception and action continuously interact to shape the agent’s experience.

\subsubsection{Internal Environment}

Although sensory information originates externally, its interpretation is fundamentally shaped by the organism's internal state. In biological systems, the brain does not assign fixed meaning to environmental stimuli; rather, it continuously evaluates them against ongoing physiological needs, so that the same cue can carry vastly different motivational weight depending on the current internal condition\cite{juechems2019does}. This process is anchored in interoception, the neural mapping of bodily signals such as energy depletion, which provides the foundational input for all subsequent value computation and behavioral prioritization\cite{Craig2003}. Crucially, motivation itself is not a single signal but emerges from the interplay of two distinct processes: the encoding of internal need states, which sets the overall level of behavioral urgency, and the affective weighting of external stimuli, which determines which actions are pursued to satisfy those needs. These two processes are mirrored in the central block of Fig.~\ref{fig:framework} through two cooperating components: 

\begin{itemize}
    \item {\it Motivational Gate}: evaluates internal variables such as battery level and mission time (namely, {\em internal state}). As resources deviate from their optimal operating range, the gate generates a motivational signal whose magnitude reflects the degree of internal deficit. This signal modulates reward sensitivity by introducing a state-dependent cost term, effectively implementing a closed-loop negative feedback mechanism. Increased resource depletion elevates the penalty term, suppressing inefficient behaviors and biasing the agent toward resource-efficient trajectories. As behavior becomes more efficient and resource consumption stabilizes, the motivational signal correspondingly decreases, thereby completing the regulatory loop. %The resulting motivational signal modulates reward sensitivity based on behavioral urgency (e.g., a low battery level). 
    A mathematical formulation is provided in Sec.~\ref{sec:motivation}. 
    \item {\it Affective Representation}: assigns motivational significance to sensory inputs, enabling the agent to interpret specific environmental conditions as internally rewarding or punishing depending on the executed action. While the ``Motivational Gate" encodes urgency, the ``Affective Representation" determines directional value assignment to external cues.  The reward model is detailed in Sec.~\ref{subsec:rewards}. 
\end{itemize}

Together, these elements generate intrinsic reward-like signals that influence both \ac{MF} and \ac{MB} learning systems. Importantly, because the motivational modulation operates through bounded additive cost adjustment rather than multiplicative reward amplification, the resulting temporal-difference errors remain stable. This ensures that motivational urgency regulates behavior without destabilizing value learning, mirroring biological homeostatic control systems in which internal deficits dynamically constrain action selection. 

\subsubsection{\ac{PIT} Model-Free Learning}

The \ac{PIT} \ac{MF} module (orange block in Fig.~\ref{fig:framework}) integrates two subsystems:

\paragraph*{Pavlovian Module}
Learns predictive associations between contextual cues and motivationally significant outcomes, following approaches as in \cite{Saeidi:C26}. Its output is a state-dependent Pavlovian value that biases action selection toward approach or avoidance {tendencies and/or action invigoration or suppression}. The learning mechanism is described in Sec.~\ref{subsec:critics}.

\paragraph*{Instrumental Model-Free RL}
Learns action-outcome associations through experience, producing habitual action values that guide fast, reactive behavior. The {instrumental learning} is described in Sec.~\ref{subsec:critics}. This subsystem generates a \ac{RPE} used to assess reliability for arbitration, described in Sec.~\ref{sec:arbitration}.

\subsubsection{Model-Based Learning}
The \ac{MB} learner (yellow block in Fig.~\ref{fig:framework}) constructs and updates an internal transition model derived from experience. Using this model, the agent performs planning via simulated rollouts (Dyna-$Q$), enabling planning and long-term reasoning, and computes the corresponding model-based $Q$-values used for action evaluation. The \ac{MB} update rules and planning cycle are detailed in Sec.~\ref{sec: hybrid}. This subsystem also computes a \ac{SPE} used for arbitration as described in Sec.~\ref{sec:arbitration}.

\subsubsection{Arbitration Mechanism}

The Bayesian arbitration mechanism (green block in Fig.~\ref{fig:framework}) evaluates the relative reliability of the \ac{MF} and \ac{MB} subsystems based on their respective prediction errors. The resulting arbitration probability determines the contribution of each subsystem to the final value estimate. The arbitration process is formally described in Sec.~\ref{sec:arbitration}.

\subsubsection{Policy Design and Action Selection}

The final decision-making stage combines \ac{MF} and \ac{MB} value estimates into a hybrid action-value representation, weighted by the arbitration probability, and biased by Pavlovian values. The resulting logits drive a softmax policy that selects the agent’s action. The policy evaluation and action selection mechanism is detailed in Secs.~\ref{subsec:policy} and~\ref{sec:arbitration}.

Overall, this architecture enables the agent to integrate cue-driven responses, habitual control, and planning, achieving robust decision-making in uncertain and complex environments.

\begin{figure}[!t]
    \centering
    \includegraphics[width=1\linewidth]{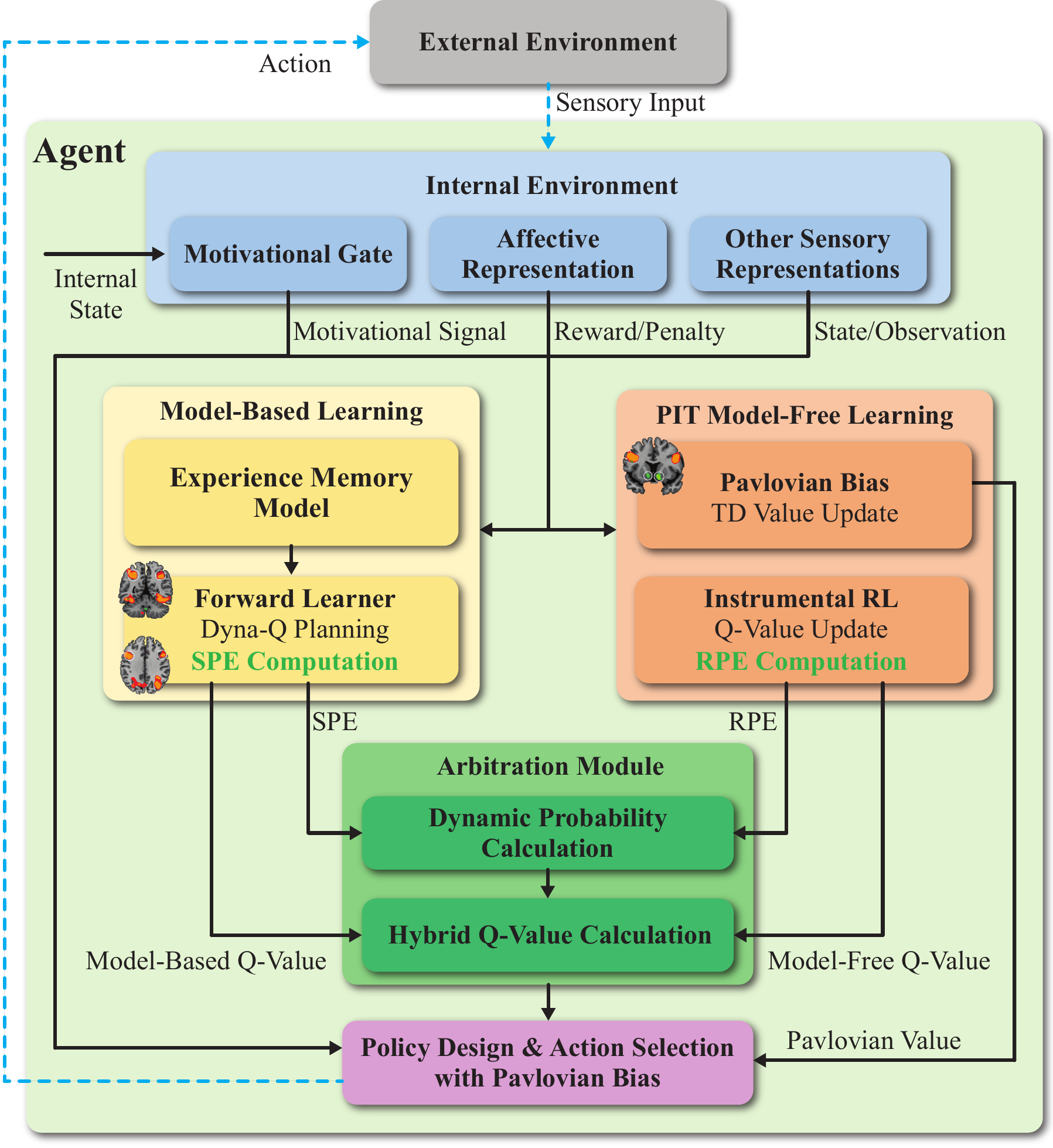}
    \caption{Introduced hybrid learning architecture combining \ac{MB} and \ac{MF} processes, inspired by neuroscience \cite{lee2019decision}. 
    }
    \label{fig:framework}
    \vspace{-10 pt}
\end{figure}

\section{Pavlovian–Instrumental Transfer (PIT) Model-Free (MF) RL Framework}
\label{sec:methodology}

We now introduce the proposed {hybrid architecture incorporating \ac{PIT} for} target 
localization, illustrated in Fig.~\ref{fig:framework}.
\subsection{Environment and Agent State Representation}
 We consider 
$N$ autonomous agents  %$\mathcal{N} = \{1, \dots, i, \dots, N\}$ 
operating in an 
unknown environment. Their objective is to navigate the area and search for a 
hidden target located at $\mathbf{p}_{\text{T}}$ by guaranteeing a localization error, in terms of \ac{PEB}, below a desired threshold. 

The environment is modeled as a discrete 2D grid with cell size $\Delta$, 
where each cell may be free, occupied by walls, agents, or the target, 
or correspond to contextual cues. Each agent learns independently and 
maintains its own state--action value function through an individual 
$Q$-table, i.e., a tabular representation that stores the estimated value 
$Q(s,a)$ associated with taking action $a$ in state $s$. This design follows the independent learners paradigm commonly adopted in  multi-agent reinforcement learning, where each agent updates its policy based on local experience while interacting with other agents through the shared environment. As a result, coordination may emerge implicitly through the environment dynamics rather than through explicit information sharing.\footnote{It is worth noting that, under the \ac{MB} approach, each agent maintains and updates an independent $Q$-table alongside its \ac{MF} counterpart.} 
The state of agent $i$ at time $t$ is given by its current position, 
$\mathbf{s}_{i,t} = \mathbf{p}_{i,t} \in \mathcal{S}$ with $i \in \mathcal{N}=\left\{1, 2, \ldots, N \right\}$. Here, $\mathcal{S}$ denotes the state space, i.e., the set of all possible positions an agent can occupy. At each step, the agent 
selects an action 
$\mathbf{a}_{i,t} \in \mathcal{A} = 
\{\text{up},\, \text{down},\, \text{right},\, \text{left},\, \text{hover}\}$, each represented as a
displacement vector of magnitude $\Delta$ (e.g., $\text{left} = [-\Delta,\, 0]$).

In addition to its external state, each agent maintains an internal state 
vector $\boldsymbol{\iota}_{i,t} = [\,b_{i,t},\ \tau_{i,t}\,]^T$,
where $b_{i,t}$ denotes the remaining battery level and $\tau_{i,t}$ the 
elapsed mission time. Executing the \textit{hover} action consumes less energy 
than movement actions, reflecting its lower operational cost. Agents are assumed to be equipped with RF sensing devices that allow them to observe electromagnetic features of the environment and exploit them as contextual cues for decision making.

\subsection{Contextual Radio Cues}
As previously discussed, \emph{potential cues} are initially \ac{NS} that  
become, after association, predictive of an outcome. 
In this regard, let here define $\mathcal{G} = \left\{\mathcal{G}_{\text{gate}} \cup \mathcal{G}_{\text{GD}}  \right\}=  \{1, \ldots, g, \ldots G\}$ as the set of indexes referring to environmental states 
representing \emph{potential cues}. Cues are here divided into two categories:  
(i) \emph{gates} ($g \in \mathcal{G}_{\text{gate}}$), namely features extracted from the \ac{EM} map corresponding 
to openings (doors or windows)\footnote{In this work, we assume that a preliminary phase that comprises the reconstruction of an electromagnetic map of the environment and the extraction of semantic radio features (e.g., the shape of a door) has already been completed.} that may indicate a transition from \ac{NLOS} 
to \ac{LOS} propagation conditions;  
(ii) \emph{GPS-denied regions} ($g \in \mathcal{G}_{\text{GD}}$), where agents lose access to absolute 
positioning, leading to increased self-localization uncertainty, which in turn degrades target-localization performance. 

Initially perceived as \ac{NS}, these potential cues become \ac{CS} 
through repeated association with favorable outcomes, with the target state 
functioning as the \ac{US}.\footnote{Throughout this 
work, the terms \emph{potential cue} and \emph{neutral stimulus} are used 
interchangeably; similarly, \emph{validated cue} and \ac{CS} are considered 
equivalent.} Table \ref{tab:pavlovian_parallelism} summarizes the parallelism between Pavlovian learning in neuroscience for humans and our interpretation in this paper for autonomous agents.
For instance, only those gates that reliably lead to advantageous areas 
(e.g., \ac{LOS} regions facilitating target discovery) are validated as 
\ac{CS}s, whereas gates that do not improve target localization remain 
behaviorally irrelevant. In the remainder of this work, we assume that the 
association between \ac{CS} and outcome has already been established. A 
possible algorithm to implement such cue–outcome association 
is discussed in~\cite{Saeidi:C26}.

\begin{table*}[t!]
\centering
\caption{Parallelism between Pavlovian conditioning, and our proposed framework.}
\label{tab:pavlovian_parallelism}
\renewcommand{\arraystretch}{1.2}
\begin{tabularx}{\textwidth}{l X X}
\hline
\textbf{Pavlovian learning (neuroscience)} &
\textbf{Description} &
\textbf{In this paper} \\
\hline
Neutral Stimulus (NS) &
State with no intrinsic value, that is, before association they are not linked with task outcomes. &

$(i)$ GPS-denied regions 

$(ii)$ Gates 

These contextual radio features, {\emph{before association}}, are not initially associated with task outcomes.

\\ \hline

Unconditioned Stimulus (US) &
Task-relevant outcome signal. &
Target localization metrics, such as the position error bound (PEB), depend on various factors. Among them, we focus on:

$(i)$ Agents' position accuracy,

$(ii)$ \ac{LOS} propagation conditions.

 \\ \hline
Conditioned Stimulus (CS) &
Predictive state acquired through learning. &

When GPS-denied regions and gates are considered validated contextual radio-cues predictive of outcomes learned from association:

$(i)$ GPS-denied regions anticipate uncertainty on agents' positions and, thus, decreased \ac{PEB}

$(ii)$ Gates anticipate LOS propagation and, thus, improved \ac{PEB}

 \\ \hline

Unconditioned Response (UR) &
Immediate reflexive response to reinforcement. &
Instantaneous value update driven by reward-based temporal-difference error. \\ \hline

Conditioned Response (CR) &
Automatic, stimulus-driven behavioral bias. &
$(i)$ GPS-denied region: avoidance tendency induced by Pavlovian values through action-dependent policy modulation.

$(ii)$ Gates: approach tendency induced by Pavlovian values through action-dependent policy modulation.\\
\hline
\end{tabularx}
\end{table*}

\subsection{Motivational Gate and Internal Drive Modulation}
\label{sec:motivation}
To mimic how internal motivational drives influence biological behavior, we define a motivational signal $\Mit$ for autonomous agents as
\begin{equation}\label{eq:motivation}
	\Mit = \xi_1(1- \bit) + \xi_2 \cdot \tauit,
\end{equation}
where 
$$\bit= \frac{b_{i,t}}{B_{\text{max}}} \in [0, 1], \quad \quad \tauit = \frac{\tau_{i,t}}{T_{\text{max}}}\in [0, 1]$$ 
denote the normalized battery level and elapsed mission time, respectively. $B_{\text{max}}$ and $T_{\text{max}}$ refer to the maximum battery level and available mission time, while $\xi_1$ and $\xi_2$ represent the weights for the battery and mission time components, respectively. In this work, these parameters are treated as design parameters and selected empirically, the values used in the experiments are reported in Table~\ref{tab:hyperparameters}.

This motivational mechanism is inspired by biological principles in humans, where motivation arises from metabolic needs (e.g., hunger or thirst), emotional states, and goal-directed behaviors. At the neural level, interoceptive signals, originating from visceral, metabolic, and somatic receptors, are continuously mapped by the insular cortex into structured representations of physiological state, providing the foundational input for all subsequent motivational and value computations\cite{Craig2003}. These signals are relayed along a posterior–anterior gradient through brainstem nuclei, thalamus, and insula, before reaching higher-order regions such as the anterior cingulate cortex (ACC) and ventromedial prefrontal cortex (vmPFC)\cite{Boorman2013}, where they are progressively integrated with affective, motivational, and cognitive information.

Homeostatic deficits, such as low energy reserves or time pressure, are translated into directed motivational drives primarily via the lateral hypothalamus (LH), which encodes motivational salience in a state-dependent manner, sharpening the distinction between appetitive and aversive inputs as need states intensify\cite{Torii1998, Ono1986}. Full motivational activation emerges through the mesolimbic dopamine system, especially the projections from the ventral tegmental area (VTA) to the nucleus accumbens, which attribute incentive salience, or “wanting,” to cues and actions relevant to current needs \cite{Panksepp2008}. Crucially, this cue-dependent mechanism is complemented by a hypothalamus–ventral striatum–VTA seeking circuit capable of independently energizing exploratory behavior even in the absence of explicit external cues, suggesting that internal need states alone carry sufficient motivational weight to initiate action\cite{Bosulu2022}.

Beyond energizing behavior, these biological systems operate as regulatory feedback loops: as physiological resources deviate from their optimal range, deficit signals intensify, increasing motivational urgency and biasing behavior toward resource-restoring actions\cite{Cisek2019}. Once homeostasis is re-established, the urgency signal diminishes, stabilizing behavior. Motivation thus functions not merely as amplification, but as dynamic negative feedback control over action selection and value computation. 

Drawing on this analogy, the motivational signal in \eqref{eq:motivation} allows each autonomous agent to adjust its learning dynamics and decision-making according to its internal state, thus capturing a simplified affect-like modulation \cite{dayan2002reward, Salichs2012}.  The two components of $\Mit$ correspond to distinct yet complementary biological drives. The battery depletion term $\bit= \frac{b_{i,t}}{B_{\text{max}}} \in [0, 1]$ reflects interoceptive deficit signals, similar to hunger or energetic need, that increase motivational urgency as resources decline, while the elapsed mission time term $\tauit = \frac{\tau_{i,t}}{T_{\text{max}}}\in [0, 1]$ captures a goal-directed temporal pressure, analogous to the increasing urgency to find food the longer an organism has gone without eating. 

Within the proposed architecture, these signals modulate value computation through an additive cost term, implementing a closed-loop negative feedback mechanism\cite{Cisek2019}. As internal deficits increase, the motivational penalty strengthens, suppressing inefficient or resource-intensive behaviors. The resulting policy adjustment reduces further depletion, which in turn attenuates the motivational signal. Through this recursive interaction, the agent’s behavior becomes progressively more resource-efficient under constraint, while maintaining stable learning dynamics.

\subsection{Pavlovian and Instrumental Learning}

\label{subsec:critics}
In our framework, both the instrumental and Pavlovian systems rely on value-based \ac{RL}. Formally, following the standard definition by \cite{sutton1998reinforcement}, the value of a generic state $s_t=s$ under policy $\pi$ is defined as
\begin{equation}\label{eq:valuedef}
    V_{\pi}(s) = \mathbb{E}_{\pi} \left[ \sum_{k=0}^{\infty} \gamma^k r_{t+k+1} | s_t = s\right],
\end{equation}
where $s_t$ denotes the state observed by a generic agent at time $t$. For notational simplicity, the agent index $i$ used elsewhere in the paper is omitted in this definition. $r_{t+1}$ is the reward received after transitioning from $s_t$ to the future state $s_{t+1}$, $\gamma$ is the discount factor ($0 \leq \gamma \leq 1$), determining the weight of future rewards relative to immediate ones. Similarly, the action-value function is expressed as
\begin{equation}\label{eq:qvaluedef}
    Q_{\pi}(s,a) = \mathbb{E}_{\pi} \left[ \sum_{k=0}^{\infty} \gamma^k r_{t+k+1} | s_t = s, a_t = a\right].
\end{equation}
Note that the expectation in \eqref{eq:valuedef}-\eqref{eq:qvaluedef} is taken over the stochastic trajectories induced by the environment dynamics and by the policy itself.
Building on these definitions, our proposed framework in Fig.~\ref{fig:framework} decomposes \ac{RL} into two interacting systems: an instrumental system that learns action-outcome contingencies through trial-and-error, and a Pavlovian system that predicts motivationally significant outcomes based on contextual radio cues.

The instrumental \ac{MF} system employs {off-policy} \ac{TD} learning ($Q$-learning) to estimate the state–action value function $Q_{\text{MF},i}{(\mathbf{s}_{i,t},\mathbf{a}_{i,t}) }$, $\forall i \in \mathcal{N}$. More specifically, for the $i$-th agent, the state-action value is updated as 
\begin{align}\label{eq:instrumentalupdate}
& Q_{\text{MF},i}(\mathbf{s}_{i,t}, \mathbf{a}_{i,t}) \leftarrow Q_{\text{MF},i}(\mathbf{s}_{i,t}, \mathbf{a}_{i,t}) + \alpha_{\text{Q}} \, \delta_{\text{Q}, i,t},
\end{align}
with $\alpha_{\text{Q}}$ being the learning rate and $\delta_{\text{Q}, i,t} $ being the TD error computed as
\begin{align}
	\delta_{\text{Q}, i,t} = & (r_{i,t+1} - \phi \Mit)  \nonumber\\
& + \gamma_{\text{Q}} \, {\max_\mathbf{a}} \,Q_{\text{MF},i}(\mathbf{s}_{i,t+1}, \mathbf{a}_{i,t+1}) - Q_{\text{MF},i}(\mathbf{s}_{i,t}, \mathbf{a}_{i,t}),
\label{eq:RPE}
\end{align}
where $\phi$ is a weight to balance the strength of motivational signal.

In parallel, the Pavlovian module learns a value function that for the $i$-th agent is given by
\begin{align}\label{eq:pavlovianupdate}
& V_i(\mathbf{s}_{i,t}) \leftarrow V_i(\mathbf{s}_{i,t}) + \alpha_{\text{V}} \, \delta_{\text{V}, i,t},
\end{align}
with $\alpha_{\text{V}}$ being the learning rate and $\delta_{\text{V}, i,t} $ being the \ac{TD} error computed as
\begin{align}
&	\delta_{\text{V}, i,t} = (r_{i,t+1} - \phi \Mit) + \gamma_{\text{V}} V_i(\mathbf{s}_{i,t+1}) - V_i(\mathbf{s}_{i,t}).
\end{align}

In our architecture, the Pavlovian learner maps contextual radio cues into 
\emph{state-dependent value estimates} $V_i(\mathbf{s}_{i,t})$, capturing the 
motivational significance value instead of motivational significance of each cue independently of the available actions. 

In this formulation, the Pavlovian component provides a cue-driven valuation of the state, while instrumental learning retains responsibility for discriminating among actions. As learning progresses, instrumental action preferences become more stable, and behavior is increasingly shaped by the Pavlovian value associated with the current context, in line with established theoretical accounts of \ac{PIT} interactions \cite{dayan2002reward}.

This mechanism parallels biological motivational systems, where subcortical 
structures such as the hypothalamus, amygdala, and ventral tegmental area
assign value to sensory cues and modulate action readiness prior to 
instrumental deliberation \cite{juechems2019does}.

\subsection{Reward Model} 
\label{subsec:rewards}

In this subsection, we describe the reward structure used for each agent.\footnote{Throughout this work, the term \emph{reward} refers to both positive rewards and negative rewards (i.e., penalties).} The instantaneous reward received by agent $i$ after executing action $\mathbf{a}_{i,t}$ in state $\mathbf{s}_{i,t}$ is defined as
\begin{equation}\label{eq:reward}
    r_{i,t+1}(\mathbf{s}_{i,t}, \mathbf{a}_{i,t}) 
    = r_{\text{inst},i,t+1} + r_{\text{pav},i,t+1},
\end{equation}
where $r_{\text{inst},i,t+1}$ and $r_{\text{pav},i,t+1}$ denote the instrumental and Pavlovian components, respectively, as detailed next.

% ----------------------------------------------------------------------
\paragraph*{Instrumental Conditioning Reward}

The instrumental reward consists of two components: (i) an RSSI-based shaped reward encouraging movement toward positions with higher detection and localization quality, and (ii) a risk penalty discouraging collisions with walls, agents, or the target. Formally,
\begin{equation}\label{eq:instrumentalreward}
    r_{\text{inst},i,t+1}(\mathbf{s}_{i,t},\mathbf{a}_{i,t}) 
    = r_{\text{RSSI},i,t+1} + r_{\text{risk},i,t+1},
\end{equation}
where $r_{\text{RSSI},i,t+1}$ is the reward derived from the received signal strength, and $r_{\text{risk},i,t+1}$ penalizes dangerous actions.

We shape the RSSI-based reward using the measured signal strength $\text{RSS}_{i,t}$ at each time step. Since higher RSSI (and, thus, higher SNR) increases the likelihood of target detection, we adopt a normalized reward defined as
\begin{align}
%r_{\text{RSSI},i,t+1}(\mathbf{s}_{i,t}, \mathbf{a}_{i,t})
%  &= \operatorname{normalized}\{\text{RSS}_{i,t+1}\} \\
%  &= 10^{\Delta P_{\text{r},i,t}[\mathrm{dB}]/10} \in [0,1],
r_{\text{RSSI},i,t+1}=\bar{P}_{\text{r},i,t}
    \label{eq:RSS_normalized}
\end{align}
where  $\bar{P}_{\text{r},i,t}=P_{\text{r},i,t}/P_{\text{r,max}}$ is the received power normalized with respect to its maximum possible value.

Notably, the received power (in [dBm]) at agent $i$ can computed via the link-budget model:
\begin{equation} \label{eq:power}
	P_{\text{r},i,t}[\mathrm{dBm}] 
	= P_{\text{t}}[\mathrm{dBm}] + G_{\text{r},i}[\mathrm{dBi}] + G_{\text{t}}[\mathrm{dBi}] 
	- L_{i,t}[\mathrm{dB}] + S_{\text{h}},
\end{equation}
where $P_{\text{t}}$ is the transmit power, $G_{\text{t}}$ and $G_{\text{r},i}$ are antenna gains, 
$S_{\text{h}} \sim \mathcal{N}(0,\sigma_s^2)$ models log-normal shadowing, and 
$L_{i,t}=PL_{i,t} + L_{\text{walls},i,t}$ accounts for propagation and wall losses. The path-loss term is
\begin{equation}
PL_{i,t}[\mathrm{dB}] 
= 20\log_{10}\left(\frac{4\pi}{\lambda_{cw}}\right)
  + 10\eta \log_{10}(d_{i,t}),
\end{equation}
with $d_{i,t}$ the distance from agent to target, $\eta$ the varied path-loss exponent depending on \ac{LOS} or \ac{NLOS},\footnote{Since the target periodically transmits a beacon, the path-loss corresponds to a one-way link from target to agent.} and $\lambda_{cw}$ the carrier wavelength.

The reference maximum received power $P_{\text{r,max}}[\mathrm{dBm}]$ in \eqref{eq:RSS_normalized} is computed assuming minimum path-loss at $d_0=1\,\mathrm{m}$ (with grid cell size $\Delta=1\,\mathrm{m}$).

\smallskip

The risk penalty reflects collision events. Let $\mathcal{O}$ be the set of cells occupied by obstacles or other agents at time $t+1$. Then,
\begin{equation}\label{eq:rewrisk}
r_{\text{risk},i,t+1} = -\lambda_{\text{col}}\, \mathbbm{1}\{\text{collision}_{i,t+1}\},
\end{equation}
with
\begin{gather}
\mathbbm{1}\{\text{collision}_{i,t+1}\} =
\begin{cases}
1 & \text{if } \mathbf{p}_{i,t+1} \in \mathcal{O}, \\[2pt]
1 & \text{if } \|\mathbf{p}_{i,t+1} - \mathbf{p}_{\text{T}}\| \leq d_{\text{safe}},\\[2pt]
0 & \text{otherwise},
\end{cases}
\end{gather}
and $\lambda_{\text{col}}$ and $d_{\text{safe}}$ are the collision penalty, and distance threshold between agents and the target, respectively. The operator $\mathbbm{1}$ denotes the indicator function, which is used to distinguish between cases. 

% ----------------------------------------------------------------------
\paragraph*{Pavlovian Reward}

The Pavlovian reward captures both positive and negative reinforcements arising from contextual radio cues, and is defined as
\begin{equation}\label{eq:pavlovianreward}
r_{\text{pav},i,t+1}(\mathbf{s}_{i,t},\mathbf{a}_{i,t})
= r_{\text{gate},i,t+1} + r_{\text{GD},i,t+1}.
\end{equation}
Here, $r_{\text{gate},i,t+1}$ provides a positive reinforcement when the next state belongs to a gate region (as gates are associated with \ac{NLOS}–\ac{LOS} transitions, where approaching from \ac{NLOS} typically improves propagation conditions), whereas $r_{\text{GD},i,t+1}$  introduces a negative penalty when entering a GPS-denied cell. Formally,
\begin{align}
r_{\text{gate},i,t+1} &=
\begin{cases}
+|r|, 
& \text{if } s_{i,t} \in \mathcal{S}_{\text{NLOS}},\; s_{i,t+1} \in \mathcal{G}_{\text{gate}},\\[4pt]
0, & \text{otherwise},
\end{cases} \\
r_{\text{GD},i,t+1} &=
\begin{cases}
-|r|, & \mathbf{s}_{i,t+1} \in \mathcal{G}_{\text{GD}}, \\[4pt]
0, & \text{otherwise},
\end{cases}
\end{align}
where $r$ is a fixed constant, and $\mathcal{S}_{\text{NLOS}}$ denotes the set of \ac{NLOS} states.

\paragraph*{Goal Reward} 

A terminal reward $R_{\text{goal}}$ is assigned to each agent upon successful mission completion. 
In our \ac{RL} formulation, an episode corresponds to a complete mission execution, starting from the initial deployment of the agents and ending when a termination condition or maximum mission time is met. 
The mission is considered accomplished when the target localization accuracy reaches a prescribed threshold. 
Localization accuracy is quantified through the \ac{PEB}, and an episode terminates whenever 
$\text{PEB} \leq \text{PEB}^{*}$.

The \ac{PEB} is defined as the square root of the trace of the \ac{CRLB} matrix, namely
\begin{equation}
    \mathsf{PEB}_t = \sqrt{\mathrm{trace}(\mathsf{CRLB}_t)}
                 = \sqrt{\mathrm{trace}(\mathbf{J}_t^{-1})},
\end{equation}
where $\mathbf{J}$ is the \ac{FIM} for the target position.\footnote{The CRLB expression 
follows from standard regularity conditions on the likelihood function-differentiability 
with respect to $\mathbf{p}_{\mathrm{T}}$, integrability of the score function, and vanishing 
boundary terms. In our simulations, if the \ac{FIM} $\mathbf{J}$ is not well-conditioned, 
we set $\mathsf{PEB} \rightarrow +\infty$.}

Assuming that each agent acquires ranging measurements, the \ac{FIM} is given by
\begin{equation}\label{eq:J}
    \mathbf{J}_{t}
    = \sum_{i=1}^{N} \mathbf{J}_{i,t}
    = \sum_{i=1}^{N} \frac{1}{\sigma_{i,t}^2} \,
      \mathbf{u}_{i,t} \mathbf{u}_{i,t}^\top
    \in \mathbb{R}^{2\times 2},
\end{equation}
where $\mathbf{u}_{i,t}$ is the unit vector from agent $i$ to the target, expressed as
\begin{equation}
    \mathbf{u}_{i,t}
    = \frac{\mathbf{p}_{\mathrm{T}} - \tilde{\mathbf{p}}_{i,t}}
           {\lVert \mathbf{p}_{\mathrm{T}} - \tilde{\mathbf{p}}_{i,t} \rVert}
    \in \mathbb{R}^{2\times 1},
\end{equation}
and $\tilde{\mathbf{p}}_{i,t} = \mathbf{p}_{i,t} +
\mathcal{N}(\mathbf{0},\,\mathbf{\Sigma}_{\mathrm{GPS},i,t})$
denotes the estimated agent position affected by GPS error with covariance matrix $\mathbf{\Sigma}_{\mathrm{GPS},i,t}$.

The total measurement variance $\sigma_{i,t}^2$ appearing in~\eqref{eq:J} is
\begin{equation}\label{eq:var}
    \sigma_{i,t}^2
    = \sigma_{\mathrm{r},i,t}^2
    + \mathbf{u}_{i,t}^{\top} \mathbf{\Sigma}_{\mathrm{GPS},i,t} \mathbf{u}_{i,t},
\end{equation}
where $\sigma_{\mathrm{r},i,t}^2$ denotes the variance of the ranging measurement, $\mathbf{\Sigma}_{\mathrm{GPS},i,t}$ is the covariance
matrix of the agent position estimate.
Moreover, since the GPS error affects the effective range measurement through the uncertainty in the agent position, the second term accounts for the projection of the position uncertainty onto the agent–target direction. 

The ranging variance is approximated as~\cite{4802191}\footnote{The expression in \eqref{eq:TOA} already converts temporal uncertainty into 
spatial uncertainty through the factor $c^2$. As a result, 
$\sigma_{\mathrm{ToA},i,t}^2$ represents a distance–variance term expressed in 
$\mathrm{m}^2$. }
\begin{equation}\label{eq:TOA}
    \sigma_{\mathrm{r},i,t}^2
    \approx
    \frac{c^2}{8\pi^2 \, \mathrm{SNR}_{i,t} \, \beta_{\mathrm{eff}}^2},
\end{equation}
with $c$ the speed of light, $\beta_{\mathrm{eff}}^2$ the effective bandwidth, 
and $\mathrm{SNR}_{i,t}$ the linear SNR at time $t$. The SNR is computed as $\mathrm{SNR}_{i,t}=P_{\mathrm{r},i,t}/P_{\mathrm{n},i}$ where $P_{\mathrm{n},i}$ is the receiver noise power at the $i$th agent.

\subsection{Policy Evaluation with Pavlovian Modulation} 
\label{subsec:policy}

A large body of computational work has examined how Pavlovian influences modulate instrumental action selection. Early models, such as \cite{guitart2012go}, considered decision problems in which agents choose between executing an action (``Go'') or withholding it (``No-Go''). In this context, a Pavlovian ``Go bias'' was introduced to model an intrinsic tendency toward action invigoration, independent of instrumental action–outcome contingencies. Action selection was typically implemented via a softmax rule augmented with an additive term encoding either a general action bias or a cue-dependent tendency to act.
Subsequent studies 
\cite{swart2017catecholaminergic} extended this framework by incorporating 
Pavlovian value predictions into the same softmax structure. These works 
highlighted a critical property: if the Pavlovian term enters identically for 
all actions, it cancels out during softmax normalization, producing no 
behavioral effect. A genuine Pavlovian influence on instrumental choice 
emerges only when the Pavlovian component is made action-dependent, for 
example by selectively enhancing “Go” actions or suppressing “No-Go” actions.

More recent models have emphasized the dynamic nature of this interaction. 
For instance, \cite{cavanagh2013frontal} showed that Pavlovian invigoration 
can be amplified or inhibited depending on neural markers of conflict, such 
as midfrontal theta power, allowing moment-to-moment modulation of cue-based 
biases. Other work has focused on arbitration mechanisms between Pavlovian 
and instrumental controllers \cite{dorfman2019controllability}. These studies 
propose that Pavlovian biases dominate when outcomes are relatively 
uncontrollable, whereas instrumental action values prevail when actions 
reliably influence rewards. This line of research supports the idea that 
Pavlovian and instrumental processes contribute jointly to decision-making, 
with their relative weight determined by environmental structure and  
uncertainty.

Motivated by this background, we define the action selection following the conventional softmax policy as 
\begin{equation}\label{eq:policy}
	\pi( \mathbf{a}_{i,t} \mid \mathbf{s}_{i,t}) = 
	\frac{\exp(\omega(\mathbf{s}_{i,t},  \mathbf{a}_{i,t}))/\kappa_{i,t} }
	{\sum_{\mathbf{a}\in\mathcal{A}}\exp(\omega(\mathbf{s}_{i,t},  \mathbf{a}))/\kappa_{i,t}}
\end{equation}
where $\kappa_{i,t}$ is the temperature parameter controlling the exploration level of the softmax (Boltzmann) policy and it is modulated by modulation gate and defined as $\kappa_{i,t} = \frac{\kappa_0}{1 + M_{i,t}}$. The function  $\omega(\mathbf{s}_{i,t}, \mathbf{a}_{i,t})$ is the action-selection score used by the softmax policy and defined as
\begin{equation}\label{eq:omega}
    \omega(\mathbf{s}_{i,t},  \mathbf{a}_{i,t}) = Q_{\text{MF},i}(\mathbf{s}_{i,t},  \mathbf{a}_{i,t}) + \beta \cdot g_a(V_i(\mathbf{s}_{i,t}), \mathbf{s}_{i,t}, \mathbf{a}_{i,t}),
\end{equation}
with $\beta$ being a fixed balancing parameter controlling the trade-off between {instrumental and Pavlovian learning}. 

In classical softmax action selection, any term that is identical across actions cancels out during normalization, thereby exerting no influence on the policy. For this reason, the Pavlovian contribution must be action-dependent in order to bias choice behavior. In \eqref{eq:omega}, this is achieved through the modulation function $g_a(\cdot)$, which assigns different values to each action based on the predicted Pavlovian value of the next state, and it is defined as
\begin{align}
g_a&(V_i(\mathbf{s}_{i,t}), \mathbf{s}_{i,t}, \mathbf{a}_{i,t}) = \nonumber \\
&
\begin{cases}
+|V_i(\mathbf{s}_{i,t})|, & 
\text{if } s_{i,t} \in \mathcal{S}_{\text{NLOS}},\; s_{i,t+1} \in \mathcal{G}_{\text{gate}}, \\[4pt]
-|V_i(\mathbf{s}_{i,t})|, & \text{if  } (\mathbf{s}_{i,t}, \mathbf{a}_{i,t}) \rightarrow   \mathbf{s}_{i,t+1} \in \mathcal{G}_{\mathrm{GD}}, \\[4pt]
0,              & \text{otherwise},
\end{cases}
\label{eq:cue_modulation}
\end{align}
where a  \emph{positive modulation} $+|V_i|$ promotes approach behavior, while   a \emph{negative modulation} $-|V_i|$ implements avoidance of low-value or  risky regions. 

This mechanism allows Pavlovian cue-based predictions to bias the softmax 
policy selectively, enhancing approach tendencies toward desirable regions 
and discouraging movement toward unsafe or undesirable zones.

Accordingly, the resulting policy in \eqref{eq:policy} implements a structured 
deviation from purely value-maximizing behavior: actions leading toward states 
with high Pavlovian value (e.g., gates) receive an additive positive shift, 
whereas actions leading toward aversive states (e.g., the GPS-denied area) are 
penalized by a negative shift.

Importantly, this additive modulation preserves the instrumental contingencies 
encoded in $Q_i(\mathbf{s}_{i,t},\mathbf{a}_{i,t})$ while enabling 
stimulus-driven influences on the action policy. As a 
result, the agent exhibits behavior analogous to \ac{PIT}, in which 
environmental cues systematically bias action selection even when the cue 
provides no information about which action is instrumentally optimal. This 
formulation enables autonomous agents to treat radio cues as predictive 
signals, supporting anticipatory navigation and faster adaptation in 
environments where specific regions reliably indicate risk or opportunity.

\begin{algorithm}[t!]
\caption{Training Loop for Target Search with \ac{PIT} \ac{MF} RL.} 
\label{alg:alg1}

\For{$ep = 1:N_{\text{episodes}}$}{
    Reset environment, agent states, and internal variables\;
    \textit{done} $\gets$ \textbf{false}\;
    
    \For{each agent $i = 1:N$}{
        Initialize {agents} state $\mathbf{s}_{i,1}$ and internal state $\boldsymbol{\iota}_{i,1}$\;
        Select initial action $\mathbf{a}_{i,1}$ using softmax policy $\pi(\mathbf{a}|\mathbf{s}_{i,1})$ using~(9)\;
    }

    \For{$t = 1:N_{\text{steps}}$}{
        \If{done}{
            \textbf{break} \tcp*{\textbf{end episode}}
        }
        
        %---------------------------
        \tcp{\textbf{Environment transition and measurements}}
        \For{each agent $i = 1:N$}{
            Apply action $\mathbf{a}_{i,t}$ and compute next state $\mathbf{s}_{i,t+1}$ (positions, collisions)\;
            {Compute received power, and Fisher Information Matrix using (16) and (25)}\;
        }
        {Compute global $\mathsf{PEB}_t$.}\;

        %---------------------------
        \tcp{\textbf{Reward and TD updates}}
        \For{each agent $i = 1:N$}{
            Compute motivational signal $M_{i,t}$ using~(1)\;
            Compute instrumental reward $r_{\text{inst},i,t+1}$ using~(13)\;
            Compute Pavlovian reward $r_{\text{pav},i,t+1}$ using~(20)\;
            Compute total reward $r_{i,t+1}$ using~(12)\;

            \If{$\mathsf{PEB}_t \le \mathsf{PEB}^*$}{
                $r_{i,t+1} \gets r_{i,t+1} + R_{\text{goal}}$\;
                \textit{done} $\gets$ \textbf{true}\;
            }

            Update Pavlovian value $V_i(\mathbf{s}_{i,t})$ using~(6)\;
            
            Update instrumental value $Q_i(\mathbf{s}_{i,t},\mathbf{a}_{i,t})$ using~(4)\;
            Select next action $\mathbf{a}_{i,t+1}$ using softmax policy $\pi(\mathbf{a}|\mathbf{s}_{i,t+1})$ using~(9)\;
        }

        %---------------------------
        Update {agents state $\mathbf{s}_{i,t+1}$ and internal state $\boldsymbol{\iota}_{i,t+1}$}\;
    }
}

\end{algorithm}

The complete Pavlovian-Instrumental RL loop, integrating the reward structure and the learning rules from the previous subsections, is summarized in Algorithm 1. At every timestep, the agent: $(i)$ executes an action drawn from the softmax policy; $(ii)$ observes the next state and evaluates instrumental and Pavlovian rewards; $(iii)$ computes the motivational signal; and $(iv)$ {updates both the instrumental $Q$-learning value function and the Pavlovian state-value function through their respective \ac{TD} errors, from which the advantage function is derived to facilitate \ac{PIT} by decomposing action-values into state baselines and action-specific deviations. The process continues until the mission is terminated in accordance with a localization accuracy criterion.

\section{Hybrid Model-Based and Model-Free Learning}
\label{sec: hybrid}

While the \ac{MF} system governs habitual, reflexive behaviors, the \ac{MB} system enables agents to reason about future states through an internal model {of the environment} \cite{sutton1998reinforcement}. Thus, differently from \ac{MF}, \ac{MB} learning allows agents to evaluate potential actions offline, reducing real-world trial-and-error and enhancing efficiency in uncertain settings. %The proposed framework is summarized in Algorithm~\ref{alg:alg2}.

In our framework, in the \ac{MB} architecture, we employ a Dyna-$Q$ approach, which integrates real experiences with planning to accelerate learning. Following a standard definition and notation \cite{sutton1998reinforcement}, each agent maintains a tabular internal model $\mathcal{M}_i$ that records observed transitions and rewards for state-action pairs   %
\begin{equation}\label{eq:generalmodel}
\mathcal{M}_i(s,a) = (s', r, t),
\end{equation}
where $s$, $s'$ are the current state and the observed next state after executing action $a$, respectively, $r$ is the received reward, and timestep $t$ records when this transition was last observed.

\subsection{Experience Memory Model}
During real-world interactions at each timestep $t$, agent $i$ at state/position $\mathbf{s}_{i,t}$ executes action $\mathbf{a}_{i,t}$ and observes the resulting position $\mathbf{s}_{i,t+1}$ and reward $r_{i,t+1}$. 

The model in \eqref{eq:generalmodel} is updated as
\begin{equation}\label{eq:model}
\mathcal{M}_i(\mathbf{s}_{i,t}, \mathbf{a}_{i,t}) \leftarrow (\mathbf{s}_{i,t+1}, r_{i,t+1}, t).
\end{equation}
This assumes deterministic transitions in our grid world environment. Additionally, the agent maintains a list of visited state-action pairs

\begin{align}
\mathcal{V}_i &= 
\left\{\, (\mathbf{s},\mathbf{a}) \in \mathcal{S}\times\mathcal{A} 
\;\big|\; (\mathbf{s},\mathbf{a})\ \text{visited by agent } i \ \right\},
\\
&=\left\{\, (\mathbf{s}_{i,t_k}, \mathbf{a}_{i,t_k}) \right\}_{k=1}^{L},
\label{eq:visitedset}
\end{align}
where $\{t_k\}_{k=1}^L$ denotes the ordered set of time indices at which  agent $i$ visited a new state–action pair during the current episode, with $L$ representing the total number of unique state–action pairs encountered.

This list is used during planning to sample previously experienced transitions. A state-action pair is added to $\mathcal{V}_i$ only if it has not been visited before in the current episode, ensuring diverse sampling during planning.

\subsection{Model-Based Planning with Dyna-Q Architecture}
\label{sec:dyna}

After each actual environment transition, that is, after the agent $i$ at state $\mathbf{s}_{i,t}$ executes an action $\mathbf{a}_{i,t}$ and observes the resulting next state $\mathbf{s}_{i,t+1}$ and reward $r_{i,t+1}$, the agent performs $K$ independent planning steps using the Dyna-$Q$ architecture. In this way, these planning steps enable offline value propagation, reducing the need for real-world trials. More specifically, for each planning step $k \in \{1, ..., K\}$, the following phases occur:

\begin{enumerate}
    \item \emph{Random Sampling}: State-action pairs $(\mathbf{s},\mathbf{a})$ are sampled uniformly at random from the set of visited pairs $\mathcal{V}_i$ to efficiently utilize the limited planning budget while ensuring diverse replay of past experiences and avoiding over-concentration on recent transitions:
    \begin{equation}
        (\mathbf{s}_{i,t},\mathbf{a}_{i,t}) \sim \mathrm{Uniform}(\mathcal{V}_i).
    \end{equation}
    This promotes broad exploration of experienced transitions without repetition in the current episode.
    
    \item \emph{Model Query}: Retrieve the stored next state and reward from the internal model
    \begin{equation}
    (\hat{\mathbf{s}}_{i,t+1}, \hat{r}_{i,t+1}) = \mathcal{M}_i(\mathbf{s}_{i,t},\mathbf{a}_{i,t}),
    \end{equation}
    where $\hat{\mathbf{s}}_{i,t+1}$ and $\hat{r}_{i,t+1}$ are the next state and reward according to the model.
    If no transition is stored (i.e., the model entry is empty at early episodes), this planning step is skipped.
    
    \item \emph{Temporal Difference Update}: Update the model-based $Q$-table, i.e.,  $Q_{\text{MB},i}(\mathbf{s}_{i,t},\mathbf{a}_{i,t})$, using \eqref{eq:RPE} and $\hat{r}_{i,t+1}$.
\end{enumerate}

\subsection{Integrated Action Selection and Policy Execution}

To dynamically arbitrate between \ac{MB} and \ac{MF} systems, we propose a hybrid approach that synthesizes the outputs of the \ac{MB} and \ac{MF} learning systems under the guidance of a Bayesian arbitration mechanism. 

The primary input for action selection is a hybrid action-value function, denoted with $Q_{\text{hybrid},i}(\mathbf{s}_{i,t},\mathbf{a}_{i,t})$, which serves as a unified estimator of the long-term utility of taking action $\mathbf{a}_{i,t}$ in state $\mathbf{s}_{i,t}$ for agent $i$ and is given by
\begin{align}\label{eq:Qhybrid}
Q_{\text{hybrid},i}(\mathbf{s}_{i,t},\mathbf{a}_{i,t}) &= (1 - P_{\text{MB},i,t}) \cdot Q_{\text{MF},i}(\mathbf{s}_{i,t},\mathbf{a}_{i,t}) \nonumber \\
&+ P_{\text{MB},i,t} \cdot Q_{\text{MB},i}(\mathbf{s}_{i,t},\mathbf{a}_{i,t}),
\end{align}
where $Q_{\text{MB},i}(\mathbf{s}_{i,t},\mathbf{a}_{i,t}) $ is the \ac{MB} $Q$-table computed using the Dyna-$Q$ architecture described in Sec.\ref{sec:dyna}, whereas the $Q_{\text{MF},i}(\mathbf{s}_{i,t},\mathbf{a}_{i,t}) $ is the \ac{MF} $Q$-table computed as in \eqref{eq:instrumentalupdate}, and $P_{\text{MB},i,t} \in [0,1]$
is a dynamically updated Bayesian arbitration probability (see also \ref{sec:arbitration}).

The use of a linear combination weighted by a probability ensures the agent's decision is robust: in states where the environment is predictable and the internal model is accurate ($P_{\text{MB},i, t} \rightarrow 1$), the agent relies heavily on foresight and planning. In novel or highly stochastic states where the model fails ($P_{\text{MB},i, t} \rightarrow 0$), the agent reverts to proven, habitual responses encoded in the MF system. This adaptive weighting is a key feature enabling the architecture to operate robustly across diverse or changing environmental conditions.

Actions are then selected using a softmax policy over the hybrid $Q$-values, modulated by Pavlovian values. The policy equation  in \eqref{eq:policy} is thus augmented using
\begin{equation}
   \!\!\!\!\!\!\omega(\mathbf{s}_{i,t}, \mathbf{a}_{i,t})\!=\!{Q_{\text{hybrid},i}(\mathbf{s}_{i,t}, \mathbf{a}_{i,t}) \!+\! \beta \!\cdot\! g_a(V_i(\mathbf{s}_{i,t}), \mathbf{s}_{i,t}, \mathbf{a}_{i,t})}.
\end{equation}

\subsection{Bayesian Arbitration Mechanism for Model Selection}
\label{sec:arbitration}

We here derive the arbitration probability $P_{\text{MB},i,t}$ used in 
\eqref{eq:Qhybrid}. 

\paragraph*{Arbitration Probability}

Following the Bayesian framework proposed in \cite{daw2005uncertainty, lee2014neural} for humans, 
the arbitration probability for autonomous agents can be defined as
\begin{equation}
P_{\text{MB}, i, t}
=
\frac{\chi_{\text{MB},i,t}}{\chi_{\text{MB},i,t} + \chi_{\text{MF},i,t} + \epsilon},
\end{equation}
where $\chi_{\text{MB},i,t}$ and $\chi_{\text{MF},i,t}$ denote the reliability scores of the 
\ac{MB} and \ac{MF} systems (evaluated in \eqref{eq:score}), respectively, and $\epsilon$ is a small regularization parameter 
to prevent division by zero. This quantity satisfies $P_{\text{MB},i,t}\in[0,1]$ and 
represents the relative confidence placed in the MB system: when $\chi_{\text{MB},i,t}\gg 
\chi_{\text{MF},i,t}$, we obtain $P_{\text{MB},i,t}\to 1$, giving dominance to MB planning; 
conversely, when the MF system is more reliable, $P_{\text{MB},i,t}\to 0$, favoring 
MF-driven control. The arbitration probability is re-computed at every time step, 
enabling fast adaptation to changes in model accuracy or environmental variability.

\paragraph*{Prediction Errors for MB and MF Systems}
\label{par:predictionerrors}

The arbitration mechanism is driven by two complementary prediction errors associated with the \ac{MB} and \ac{MF} learning systems. The \ac{MB} system relies on the \ac{SPE}, which captures errors in state-transition predictions, while the \ac{MF} system relies on the \ac{RPE}, reflecting errors in reward prediction. These two signals jointly inform the arbitration process by quantifying the reliability of planning-based versus habitual control.

For an agent executing action $\mathbf{a}_{i,t}$ from 
state $\mathbf{s}_{i,t}$ and transitioning to $\mathbf{s}_{i,t+1}$, the SPE is defined as
\begin{align}\label{eq:SPE}
&\text{SPE}(\mathbf{s}_{i,t}, \mathbf{a}_{i,t}, \mathbf{s}_{i,t+1})
=
\mathbbm{1}\!\left[\mathcal{M}(\mathbf{s}_{i,t}, \mathbf{a}_{i,t}) = \emptyset\right] \nonumber\\
&\quad + 
\mathbbm{1}\!\left[\mathcal{M}(\mathbf{s}_{i,t}, \mathbf{a}_{i,t}) \neq \emptyset\right]\,
\frac{d_{\text{M}}\big(\hat{\mathbf{s}}_{i,t+1},\mathbf{s}_{i,t+1}\big)}{d_\text{max}}
,
\end{align}
where $d_{\text{M}}(\cdot,\cdot)$ denotes the Manhattan distance between 
predicted and actual next states~\cite{russell1995modern}, $d_\text{max}$ denotes the maximum achievable Manhattan distance of the environment.\footnote{The Manhattan distance 
is chosen due to the grid-world structure, where movements occur along axis-aligned 
edges.} Furthermorew, the first indicator ensures $\text{SPE}=1$ when 
no model entry exists for $(\mathbf{s}_{i,t}, \mathbf{a}_{i,t})$. A low SPE indicates accurate MB predictions, while a high SPE suggests model 
inaccuracy or environmental stochasticity.

For the MF system, the \ac{RPE} corresponds to the \ac{TD} error 
\begin{align}\label{eq:RPE_2}
&\text{RPE}(\mathbf{s}_{i,t}, \mathbf{s}_{i,t+1})
= \delta_{\text{Q},i,t}. % \nonumber \\
%&= (1 + M_{t}) r_{t+1} 
%   + \gamma_{\text{Q}} \max_\mathbf{a} Q_\text{MF}(\mathbf{s}_{t+1}, \mathbf{a}_{t+1}) 
 %  - Q_\text{MF}(\mathbf{s}_{t}, \mathbf{a}_{t}),
\end{align}
Low absolute \ac{RPE} values indicate that the \ac{MF} system accurately predicts future  rewards, whereas large absolute values reflect substantial prediction errors due to  reward changes or insufficient learning.

By jointly analyzing SPE and RPE through Bayesian inference, the arbitration mechanism 
determines the relative reliability of MB and MF systems at each time step, enabling 
adaptive switching between deliberative planning and habitual, \ac{MF} control.

\paragraph*{Dirichlet Reliability Model}

To compute the reliability scores $\chi_{k,i,t}$ for each subsystem 
$k\in\{\text{MB},\text{MF}\}$, we maintain a Dirichlet distribution over three 
prediction-error categories. 
More specifically, for any prediction error $\text{PE}$ (either $\ac{SPE}$ or  $\ac{RPE}$), the category is assigned as \cite{daw2005uncertainty, lee2014neural}
\begin{equation}\label{eq:errorcategory}
    c(\text{PE}) =
    \begin{cases}
    0, & |\text{PE}| < \zeta \quad\,\,\,\,\, (\text{zero error}), \\[4pt]
    1, & \text{PE} < -\zeta \quad\,\,\,\, (\text{negative error}), \\[4pt]
    2, & \text{PE} > \zeta \qquad\,\,\, (\text{positive error}).
    \end{cases}
\end{equation}
where $\zeta$ denotes a symmetric  threshold around zero. 

This mapping captures whether the system’s prediction was essentially correct 
(category~0) or whether it produced significant underestimates (category~1) or 
overestimates (category~2). The threshold $\zeta$ controls sensitivity: smaller 
values make the arbitration mechanism more reactive, while larger values provide 
robustness against noise and small fluctuations.

Let $\boldsymbol{\lambda}_{k,i,t} = [\lambda_{k,i,t,0},\, \lambda_{k,i,t,1},\, \lambda_{k,i,t,2}]$ denote the vector of concentration parameters for system $k$, where each component 
$\lambda_{k,i,t,j}$ counts how many times error category $j$ has been observed.

Once a prediction error $\text{PE}_k$ is observed, 
we compute its category index $j=c(\text{PE}_k)$ and update the corresponding 
Dirichlet parameter:
\begin{equation}\label{eq:dirochletupdate}
\lambda_{k,i,t,j} \leftarrow \lambda_{k,i,t,j} + 1.
\end{equation}

This update constitutes a Bayesian filter that accumulates evidence on the frequency 
of each error type. The normalized vector
$\boldsymbol{\theta}_{k,i,t} = [\theta_{k,i,t,0},\,\theta_{k,i,t,1},\,\theta_{k,i,t,2}]$
represents the probability distribution over error categories for system $k$. Under 
the Dirichlet posterior, the mean and variance of $\theta_{k,i,t,0}$ (probability of zero 
error) are given by
\begin{equation}
\mathbb{E}[\theta_{k,i,t,0}] = \frac{\lambda_{k,i,t,0}}{\Lambda_{k,i,t}},
\text{Var}[\theta_{k,i,t,0}] = 
\frac{\lambda_{k,i,t,0} (\Lambda_{k,i,t} - \lambda_{k,i,t,0})}{\Lambda_{k,i,t}^2 (\Lambda_{k,i,t} + 1)},
\end{equation}
where $\Lambda_{k,i,t} = \sum_{j=0}^2 \lambda_{k,i,t,0}$ is the sum of counts.  

A subsystem is considered reliable when it frequently produces negligible errors 
(high $\mathbb{E}[\theta_{k,i,t,0}]$) and when such estimates are stable (low variance). 
Motivated by this intuition, the reliability score of subsystem $k$ is defined as
\begin{equation}\label{eq:score}
\chi_{k,i,t} = 
\frac{\mathbb{E}^2[\theta_{k,i,t,0}]}{\text{Var}[\theta_{k,i,t,0}]}
=
\frac{\lambda_{k,i,t,0}(\Lambda_{k,i,t} + 1)}{\Lambda_{k,i,t} - \lambda_{k,i,t,0}}.
\end{equation}
A higher value of $\chi_{k,i,t}$ indicates greater confidence in system $k$. In the considered navigation task, this reliability-based arbitration implies that, during early exploration, both \ac{MF} and \ac{MB} systems may be unreliable due to sparse experience and inaccurate models, resulting in mixed control. As experience accumulates, improved transition prediction and reduced state prediction errors are expected to increase the reliability of the \ac{MB} system, whereas convergence of reward prediction errors favors \ac{MF} control in later stages. These dynamics are validated through the case study presented in the next
section.

\section{Case Study: A Localization Problem}
\label{sec: casestudy_vector}

In this section, we present performance results to evaluate the performance of the proposed \ac{PIT} \ac{MB}/\ac{MF} learning approach. As discussed earlier, the autonomous agents are tasked with navigating an environment to search for a static target. The mission is deemed successful once the \ac{PEB} associated with the target’s location falls below a predefined threshold ($\text{PEB}^\star = 0.5\,\mathrm{m}$), indicating accurate and reliable target localization.

\subsection{Simulation Environment and Parameter Configuration}

\begin{figure}[t!]
    \centering
    \includegraphics[width=1\linewidth]{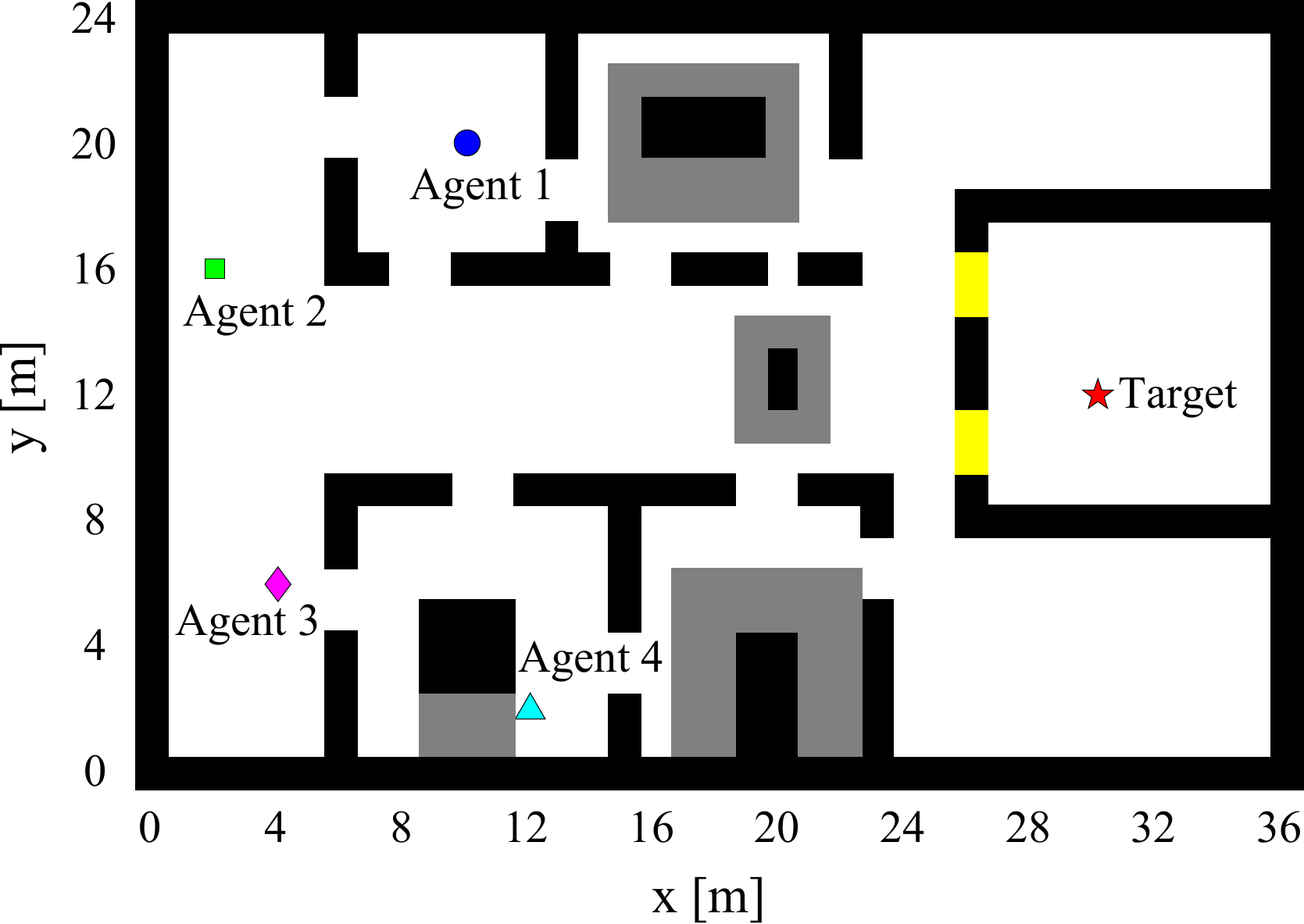}
    \caption{Simulation environment for autonomous agent network localization.}
    \label{fig:environment}
    \vspace{-20pt}
\end{figure}

The environment is modeled as a discrete 2D grid of $24 \times 36$ cells, each
measuring $\left(1 \times 1\right)\,\text{m}^2$. The layout, illustrated in
Fig.~\ref{fig:environment}, includes interior walls and obstacles (shown as
black cells), along with designated ``gate'' zones (highlighted in yellow) and
GPS-denied areas (shown in grey).

A network of $N = 4$ autonomous agents is deployed, with initial positions at positions
$\mathbf{p}_{1,0} = (10,\,20)\,\mathrm{m}$,
$\mathbf{p}_{2,0} = (2,\,16)\,\mathrm{m}$,
$\mathbf{p}_{3,0} = (4,\,6)\,\mathrm{m}$, and
$\mathbf{p}_{4,0} = (12,\,2)\,\mathrm{m}$.
A single active target, emitting periodic beacon signals, is located at position
$\mathbf{p}_{\text{T}} = (30,\,12)\,\mathrm{m}$. Each agent can select an action from a discrete set $\mathcal{A}$:
$\{\text{up}, \text{down}, \text{left}, \text{right}, \text{hover}\}$. Agent movement is constrained by the grid boundaries, walls, and other agents, with collision avoidance resolved using \eqref{eq:rewrisk}.

The reward is shaped by the \ac{RSSI} computed using \eqref{eq:power}. The path loss exponent is set to $\eta = 2$ when the agent-target link is in \ac{LOS} and to $\eta = 3.5$ otherwise. The transmitting power is set to $-10$ dBm, with both the transmitter and receiver employing antennas with 2 dBi gain. The signal occupies a 1 MHz bandwidth centered at 2.4 GHz, and the receiver noise figure is 10 dB. The wall losses $L_{\text{walls},i,t}$ are set to $25\,\text{dB}$ for all agents $i$ and time instants $t$, and only the room containing the target is considered to be in the \ac{LOS} condition. If the agent is in a GPS-denied area, its position estimate is affected by increased localization uncertainty, modeled as in \eqref{eq:var}, where the position error covariance is assumed diagonal with {$\left(\boldsymbol{\Sigma}_{\mathrm{GPS},i,t}\right)_{1,1} = \left(\boldsymbol{\Sigma}_{\mathrm{GPS},i,t}\right)_{2,2} =100\, \mathrm{m^2}$.} 

Each agent is trained with the proposed model for $N_{\text{episodes}}=1200$ episodes with a maximum step of $N_{\text{steps}}=800$. The decaying learning rates and discount factor for both the off-policy $Q$-learning and Pavlovian modulation are identical. Moreover, action selection is governed by a softmax policy with a decaying temperature $\kappa_{i,t}$. Other hyperparameters are listed in Tab. \ref{tab:hyperparameters}. 

The learning process is guided by a dual-reward system comprising instrumental and Pavlovian rewards, as described in Sec.~\ref{subsec:rewards}. In simulations, the penalty for collision in \eqref{eq:rewrisk} is set to $\lambda_{\text{col}} = 1.2$, the terminal goal reward is set to $R_{\text{goal}} = 100$, and Pavlovian rewards are defined to encode innate, cue-driven tendencies that bias the agent toward advantageous regions and away from structurally risky ones. Gates are assigned a positive Pavlovian reward because they reliably predict a rapid reduction of localization error, whereas GPS-denied regions receive negative value due to their detrimental impact on measurement precision and navigational safety. Specifically, the Pavlovian reward elements are defined as $
r_{\text{gate},i,t+1} = 8,\, \forall i,\forall t, \, r_{\text{GD},i,t+1} = -8,\, \forall i,\forall t$.
\begin{table}[!t]
        \caption{Simulation Parameters}
        \centering
        \renewcommand{\arraystretch}{1.3}
        \begin{tabular}{ccc}
            \toprule
            \textbf{Parameter} & \textbf{Description} & \textbf{Value} \\
            \midrule
            \multirow{4}{*}{ $\alpha_{\text{Q}}, \alpha_{\text{Q}_\text{MB}}, \alpha_{\text{V}}$} 
            & Learning rate: & \\
            & Initial value & 0.55 \\
            & Decay rate & 0.9985 \\
            & Final value & 0.09 \\
            \midrule
            $\gamma_{\text{Q}}, \gamma_{\text{Q}_\text{MB}}, \gamma_{\text{V}}$ & Discount factor & 0.98 \\
            \midrule
            \multirow{4}{*}{ $\kappa_0$} 
            &  Temperature: & \\
            & Initial value & 1.2 \\
            & Decay rate & 0.996 \\
            & Final value & 0.03 \\
            \midrule
            $K$ & Planning steps & 2 \\
            \midrule
            $\beta$ & \ac{PIT} balance & 1 \\
            \midrule
            $\xi_1, \xi_2$ & Motivational signal weights & 0.4, 0.4 \\ 
            \midrule
            $P_\text{MB}$ & Initial $P_\text{MB}$ & 0.5\\
            \midrule
            $\phi$ & Motivational signal balancing weight & 0.6\\
            \bottomrule
        \end{tabular}
        \label{tab:hyperparameters}
    \end{table}

\begin{figure*}[!t]
	\centering
	\subfloat[]{
		\includegraphics[width=0.32\linewidth]{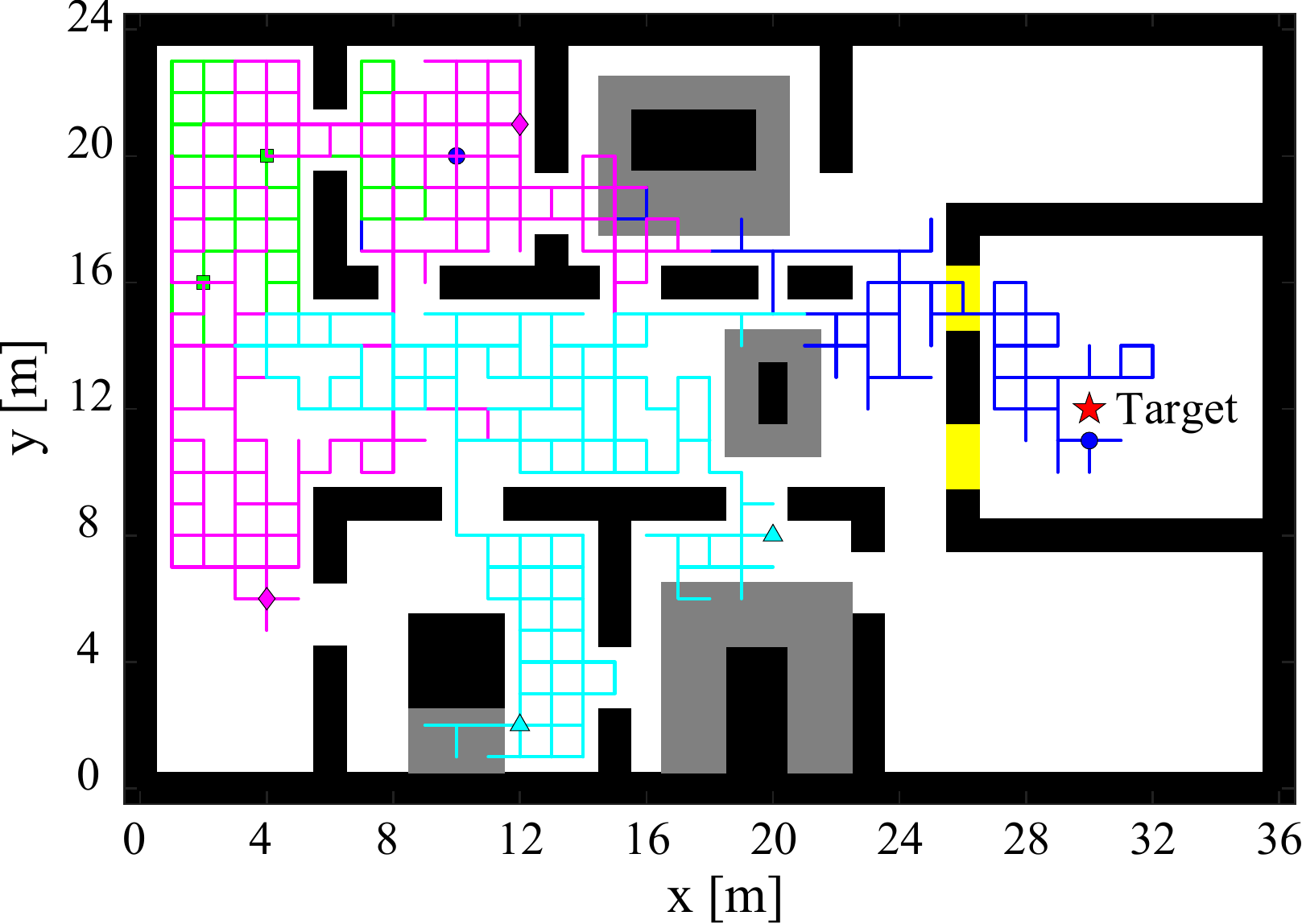}
        \label{fig:with_ep1}
	}
	\subfloat[]{
		\includegraphics[width=0.32\linewidth]{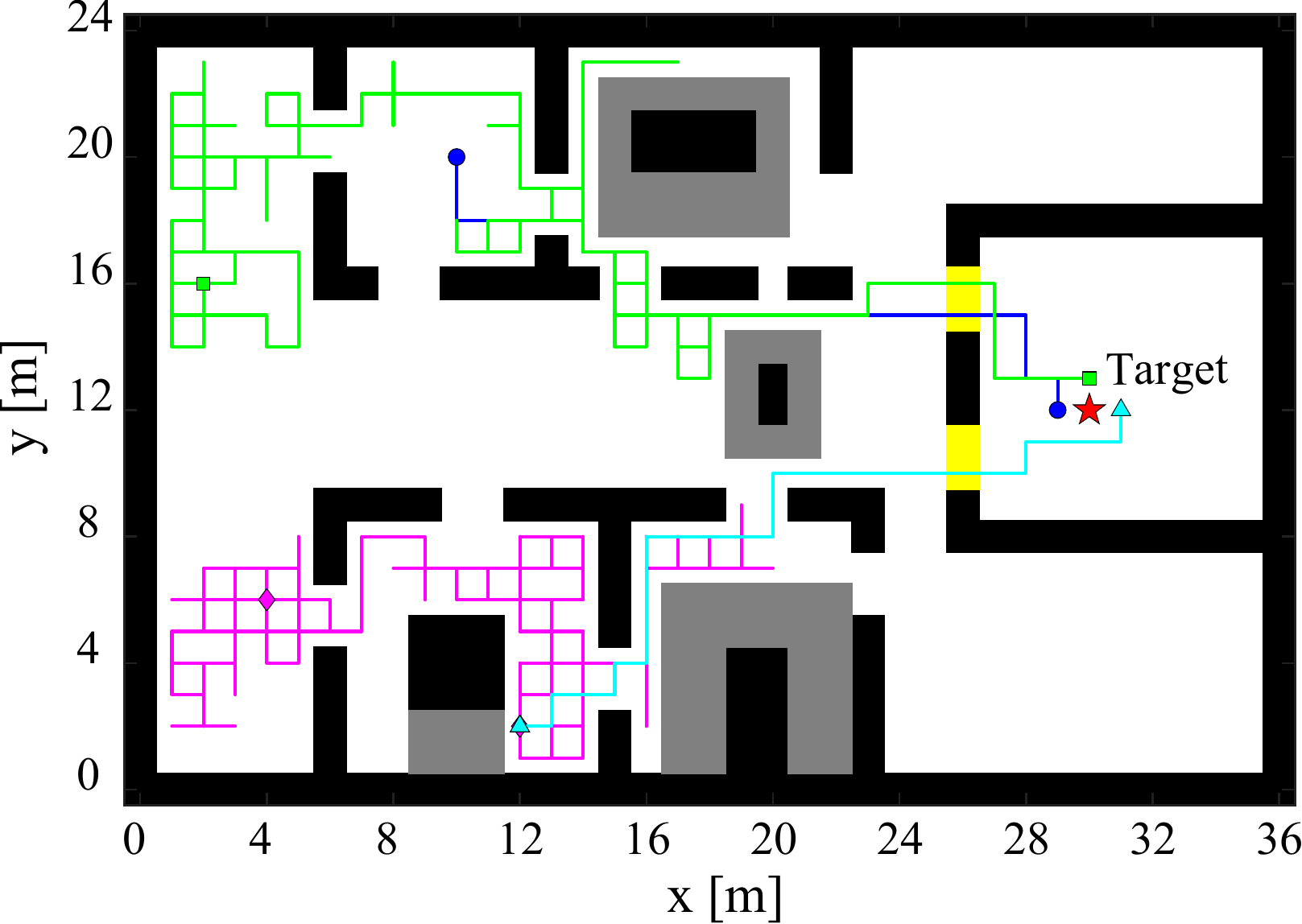}
        \label{fig:with_ep2}
	}
    \subfloat[]{
        \includegraphics[width=0.32\linewidth]{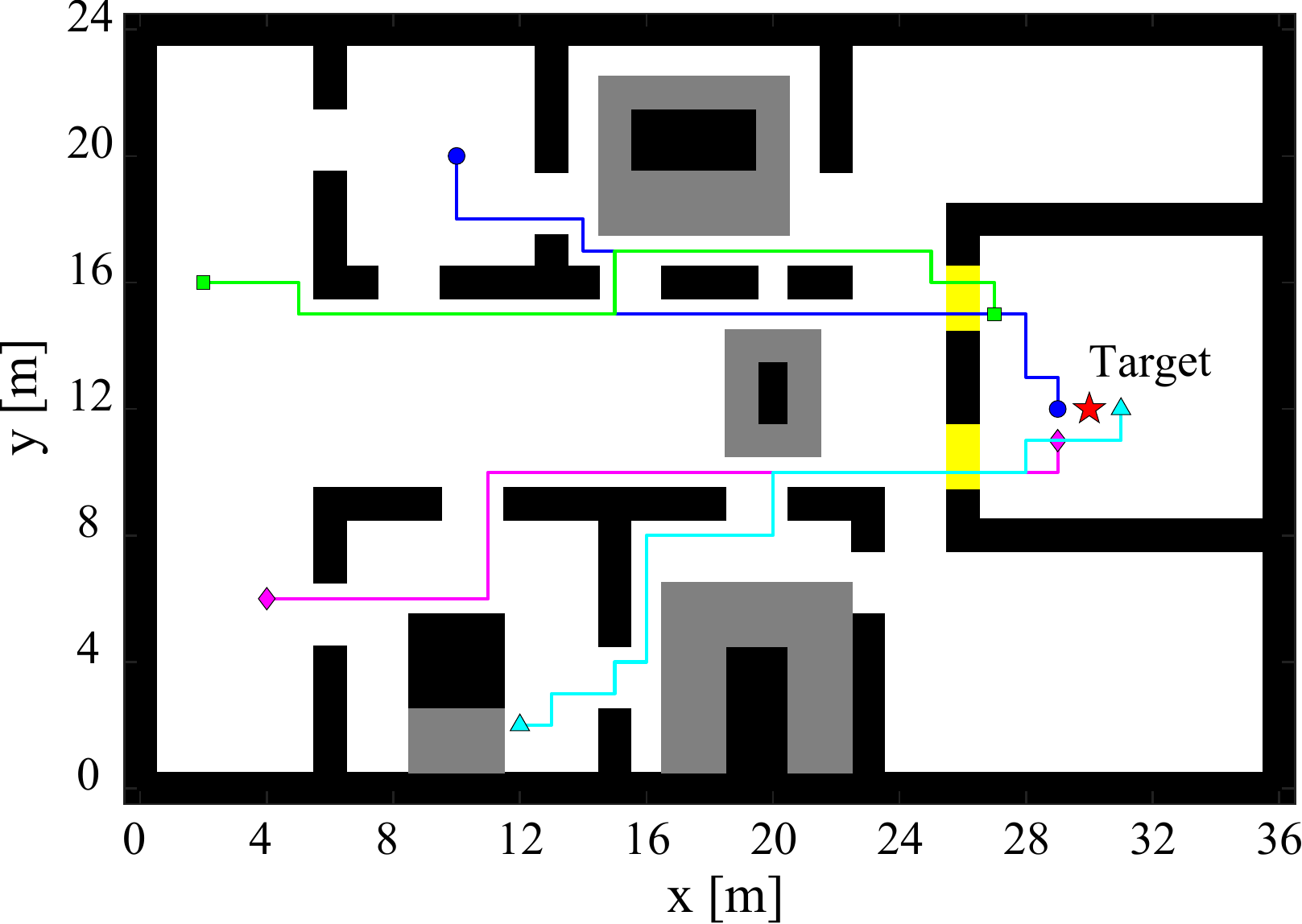}
        \label{fig:final_path}
	}
    
	\subfloat[]{
		\includegraphics[width=0.32\linewidth]{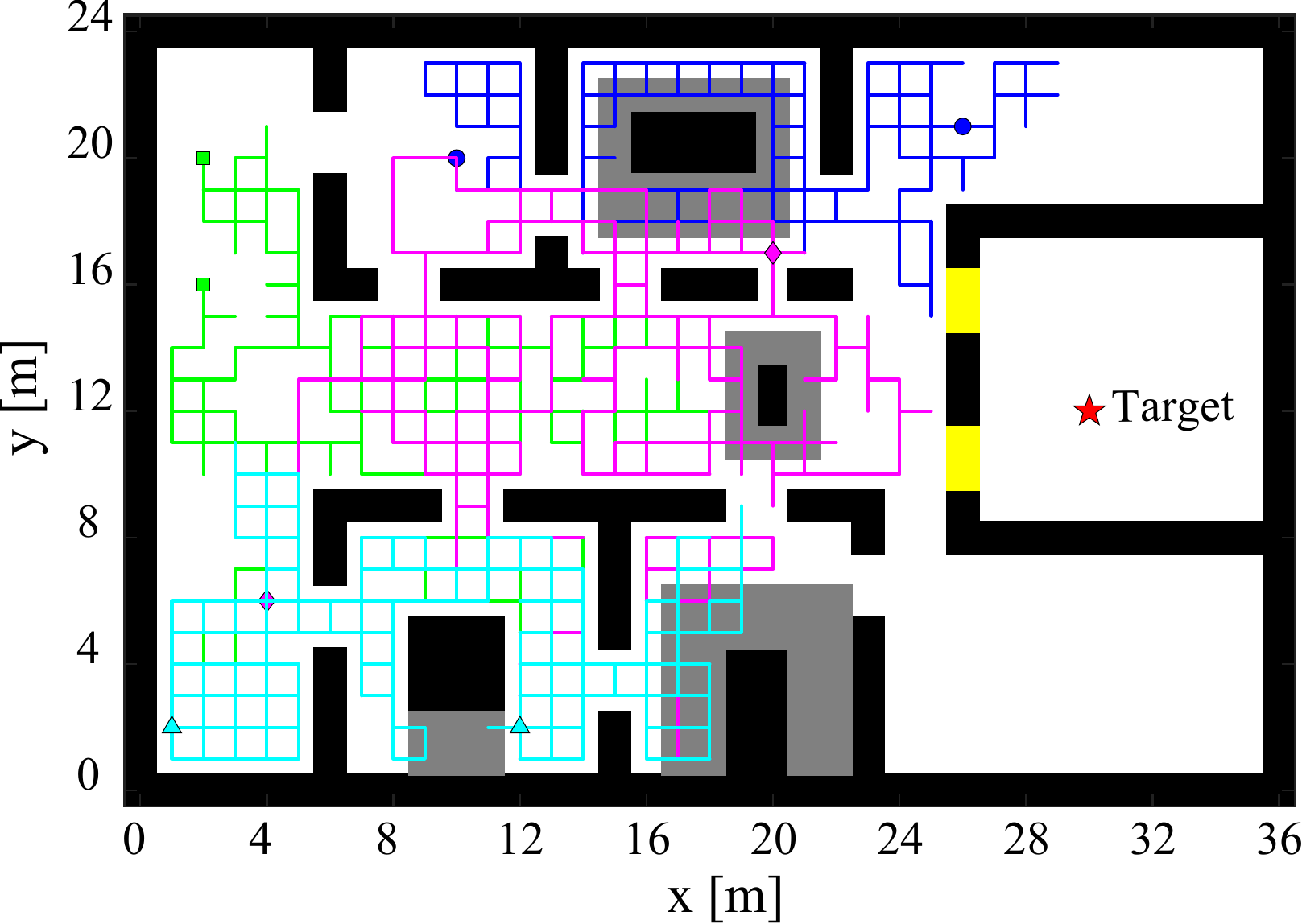}
        \label{fig:motivational_study_1}
	}
	\subfloat[]{
		\includegraphics[width=0.32\linewidth]{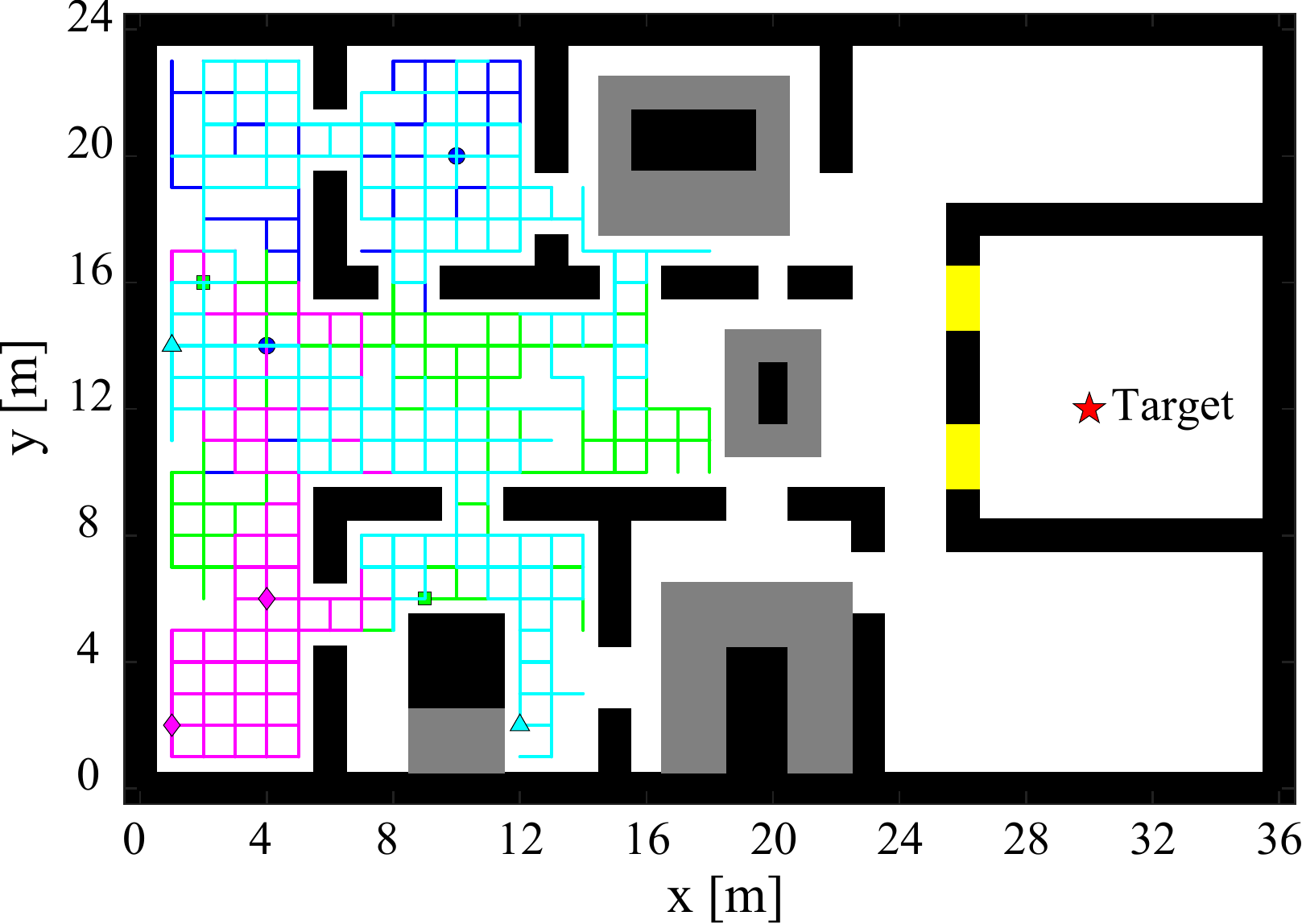}
        \label{fig:motivational_study_2}
	}
    \subfloat[]{
        \includegraphics[width=0.32\linewidth]{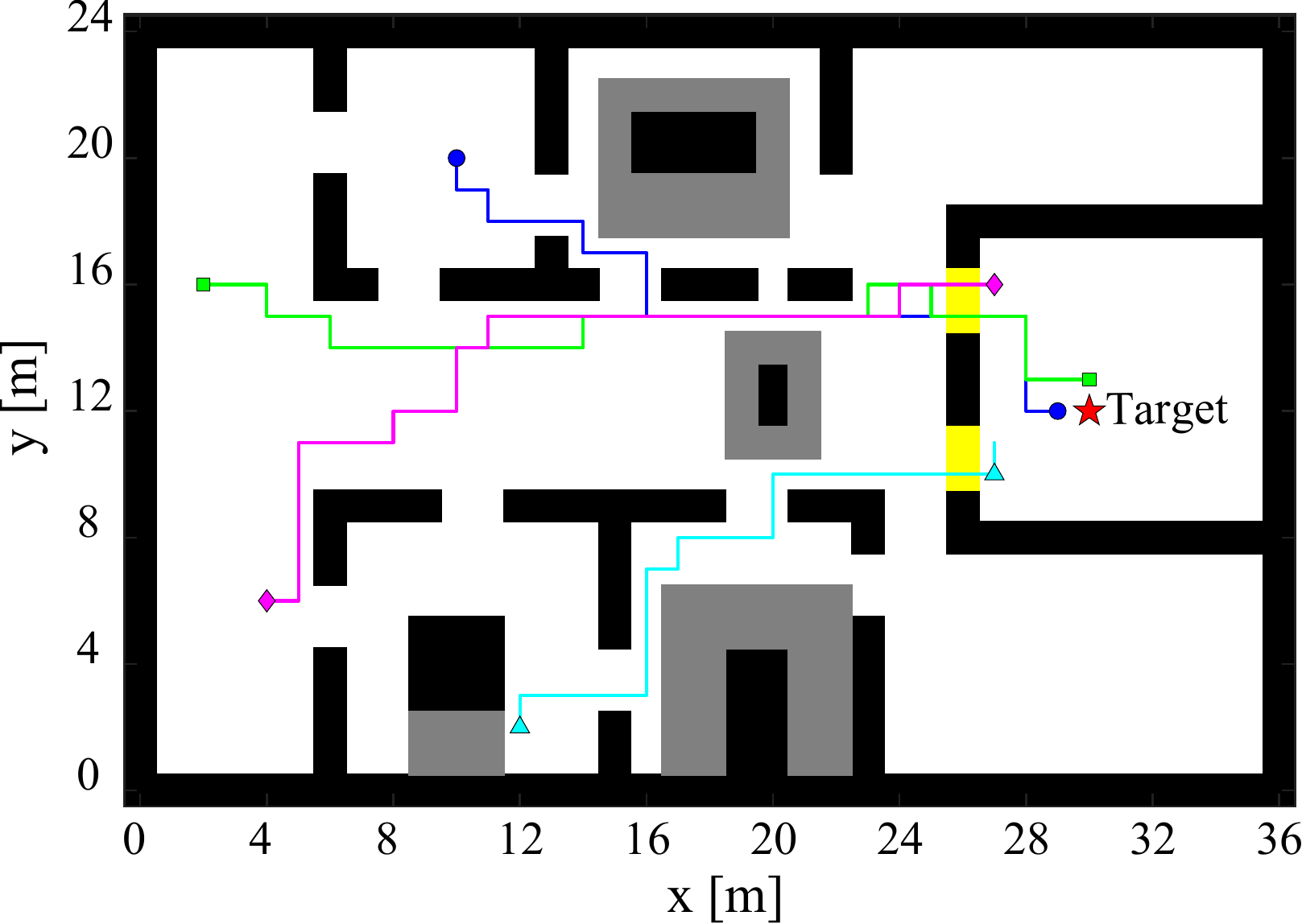}
        \label{fig:motivational_study_3}
        
	}
    
	\caption{Agent trajectories with \ac{PIT} \ac{MF} at (a) Episode 1, (b) Episode 100, (c) Episode 1200 with (top) and without (bottom) motivational signal, under the same seed. As displayed by the results, without motivational signal, agents are more incline to environmental exploration as per humans \cite{tolman1930introduction}.}	
    \label{fig:agent_trajectory_PIT}
    \vspace{-20pt}
\end{figure*}

\subsection{\ac{PIT} \ac{MF} Learning: Performance Assessment}

We first analyze Pavlovian and instrumental \ac{MF} learning in isolation to assess how cue-driven modulation shapes early exploration and affects agents’ navigation before introducing \ac{MB} planning components.

\subsubsection{Evolution of Navigation Policies Across Training}
The evolution of agent behavior is illustrated by examining their trajectories at different stages of training. Fig~\ref{fig:agent_trajectory_PIT} shows the trajectories of all four agents trained with \ac{PIT} \ac{MF} at three representative training episodes: Episode~$1$, Episode~$100$, and Episode~$1200$.

One can observe clearly that at Episode~$1$, the agents exhibit exploratory and inefficient paths, with some becoming trapped into loops or entering GPS-denied regions. This behavior reflects the high-entropy softmax policy resulting from a large temperature parameter $\kappa_{i,t}$. By Episode~$100$, the agents' trajectories become increasingly oriented toward the target. They tend to avoid GPS-denied areas and pass through gates in a more reliable manner, characterized by a higher consistency of successful gate crossings across episodes and a reduced number of failed attempts or collisions. This behavior shift reflects the influence of Pavlovian learning, which biases exploration toward contextually favorable transitions. While agents relying solely on instrumental learning are eventually able to pass through the gates, such behavior typically emerges after more prolonged exploration and exhibits greater variability across episodes. In contrast, Pavlovian cue modulation facilitates earlier and more consistent gate-crossing behavior. At this stage, agents may already reach the terminal condition, although typically via suboptimal paths.
By Episode~$1200$, the agents display highly efficient behavior. They consistently traverse the gates, maintain \ac{LOS} with the target, avoid collisions, and approach the target to minimize the \ac{PEB}, completing the task in only 31 steps. 

\subsubsection{Influence of Motivational Signal}

In addition to the Pavlovian and instrumental learning components, the influence of motivational signal $\Mit$ is also studied. The results are shown in Fig~\ref{fig:motivational_study_1}, ~\ref{fig:motivational_study_2}, and ~\ref{fig:motivational_study_3} by holding the same seed as previous simulation.

At Episode 1, agents exhibit exploratory and largely random behavior, not perform goal-directed behavior as in Fig~\ref{fig:with_ep1}. By Episode 100, a pronounced divergence emerges: agents incorporating the motivational signal begin to converge more rapidly toward regions of interest, particularly the target area, agents without motivational signal continue to display diffuse and inefficient exploration patterns. At Episode 1200, agents of both configurations consistently follow direct, collision-free trajectories toward the goal.

 These findings are consistent with seminal works in neuroscience like \cite{tolman1948cognitive, tolman1930introduction}, which demonstrated that rats allowed to explore a maze without hunger motivation exhibited diffuse exploratory behavior but later displayed rapid, goal-directed navigation once motivation was introduced. In particular, similar to the interpretation of latent learning and cognitive map formation in \cite{tolman1948cognitive} our results suggest that the motivational signal does not merely enhance reward sensitivity but reorganizes exploratory dynamics into structured, goal-pursuit strategies, thereby accelerating convergence under resource constraints. 
\vspace{-5pt}

\begin{figure*}[!t]
	\centering
	\subfloat[]{
		\includegraphics[width=0.238\linewidth]{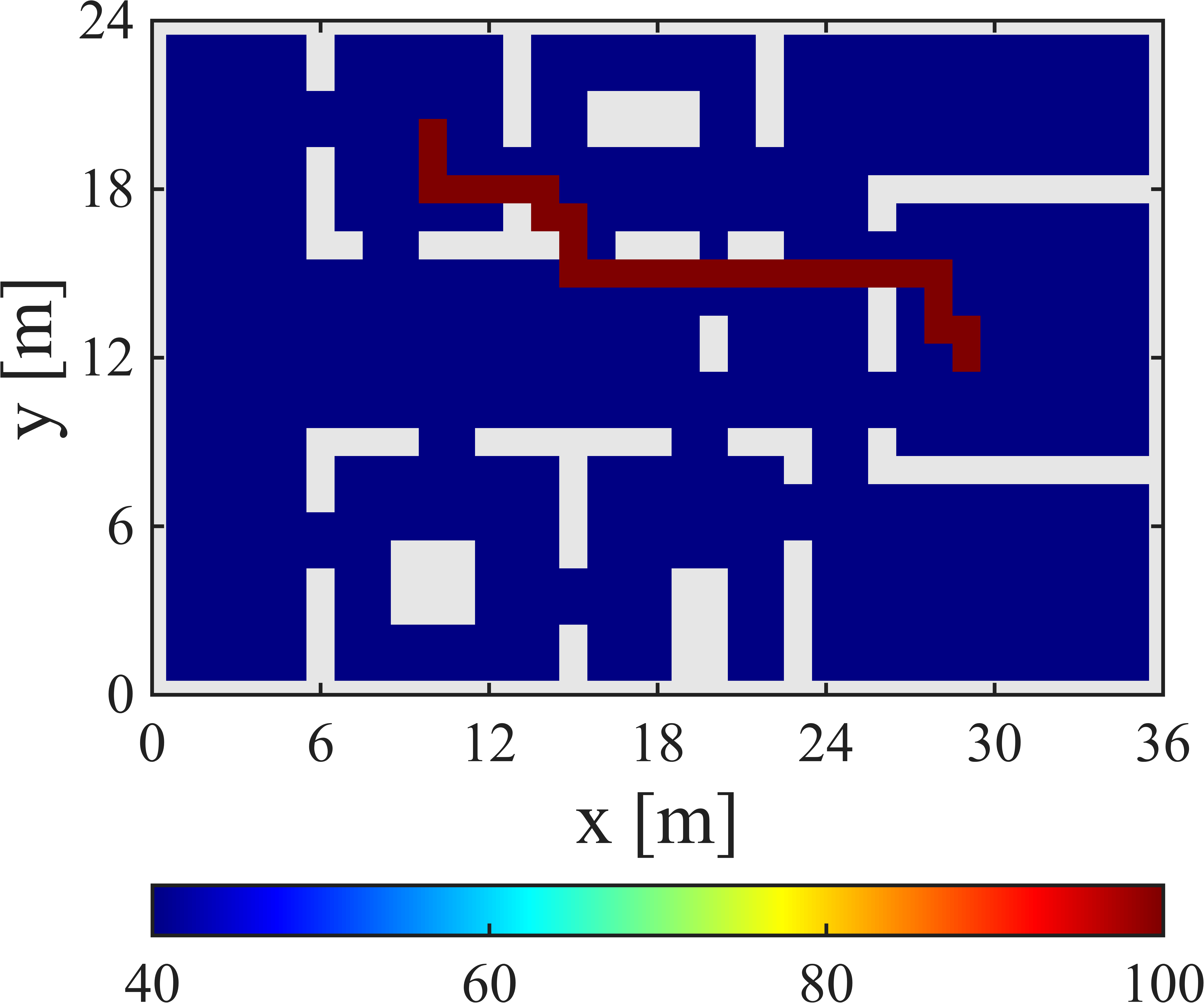}
	}
	\subfloat[]{
		\includegraphics[width=0.238\linewidth]{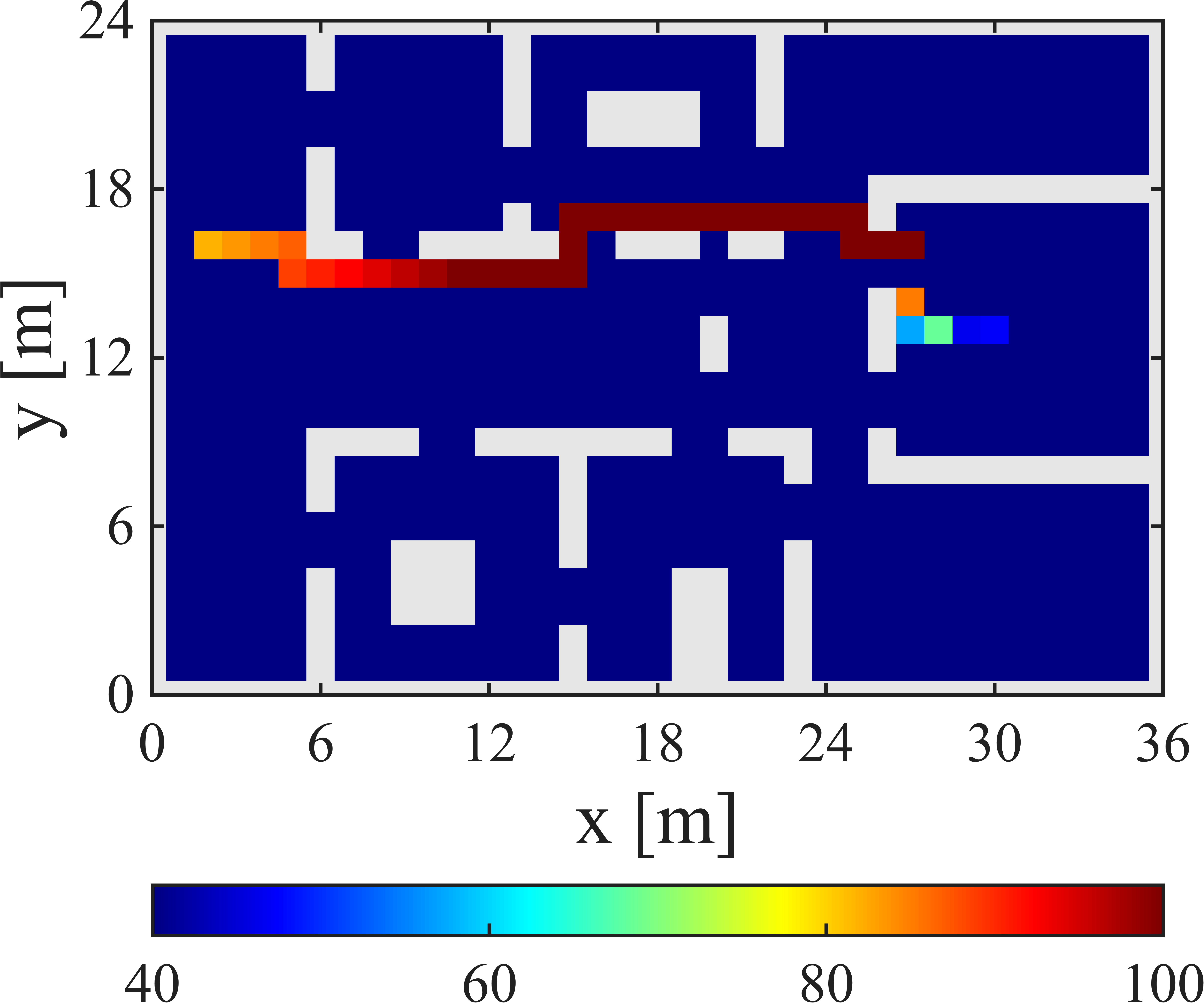}
	}
    \subfloat[]{
        \includegraphics[width=0.238\linewidth]{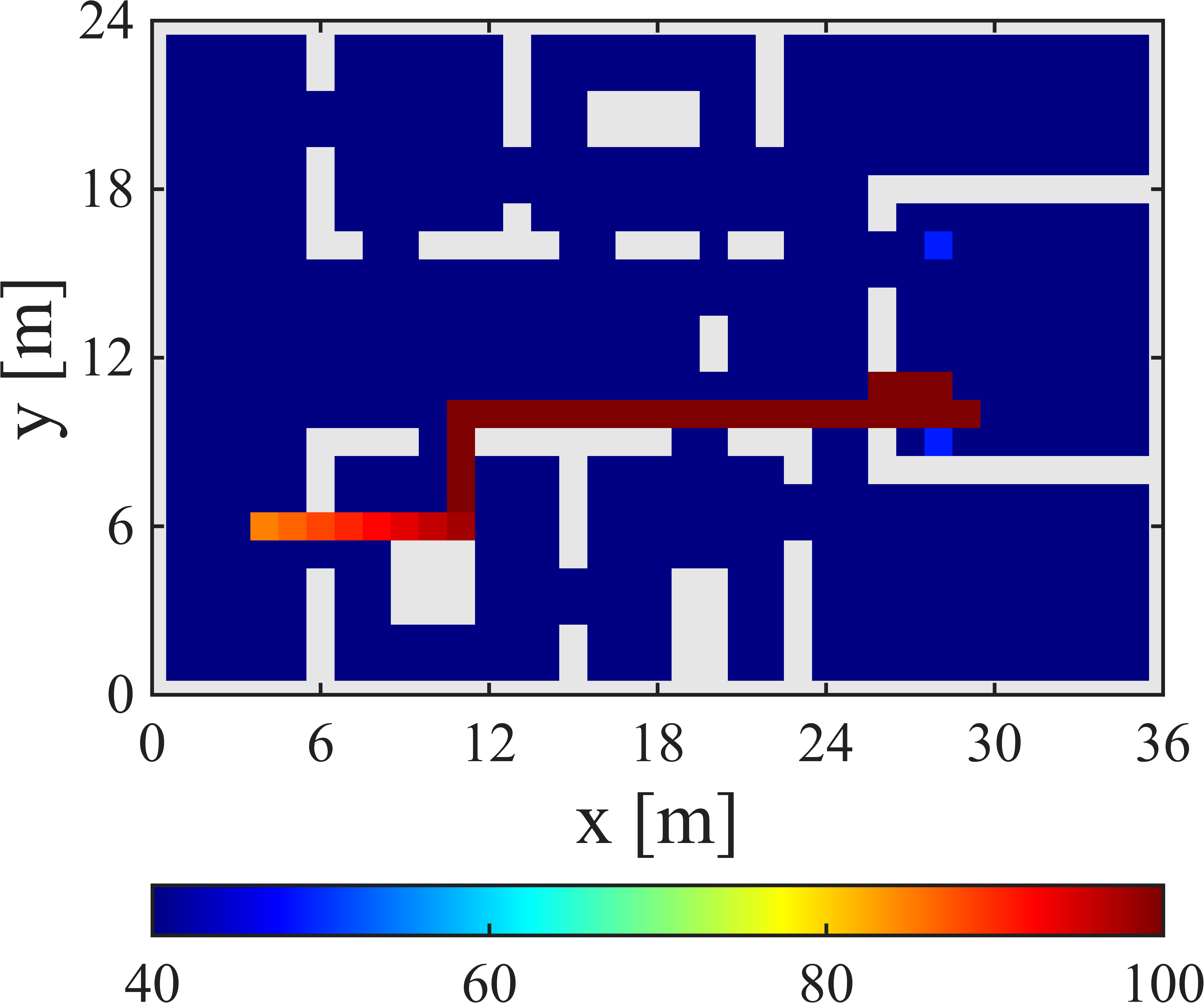}
	}
	\subfloat[]{
		\includegraphics[width=0.238\linewidth]{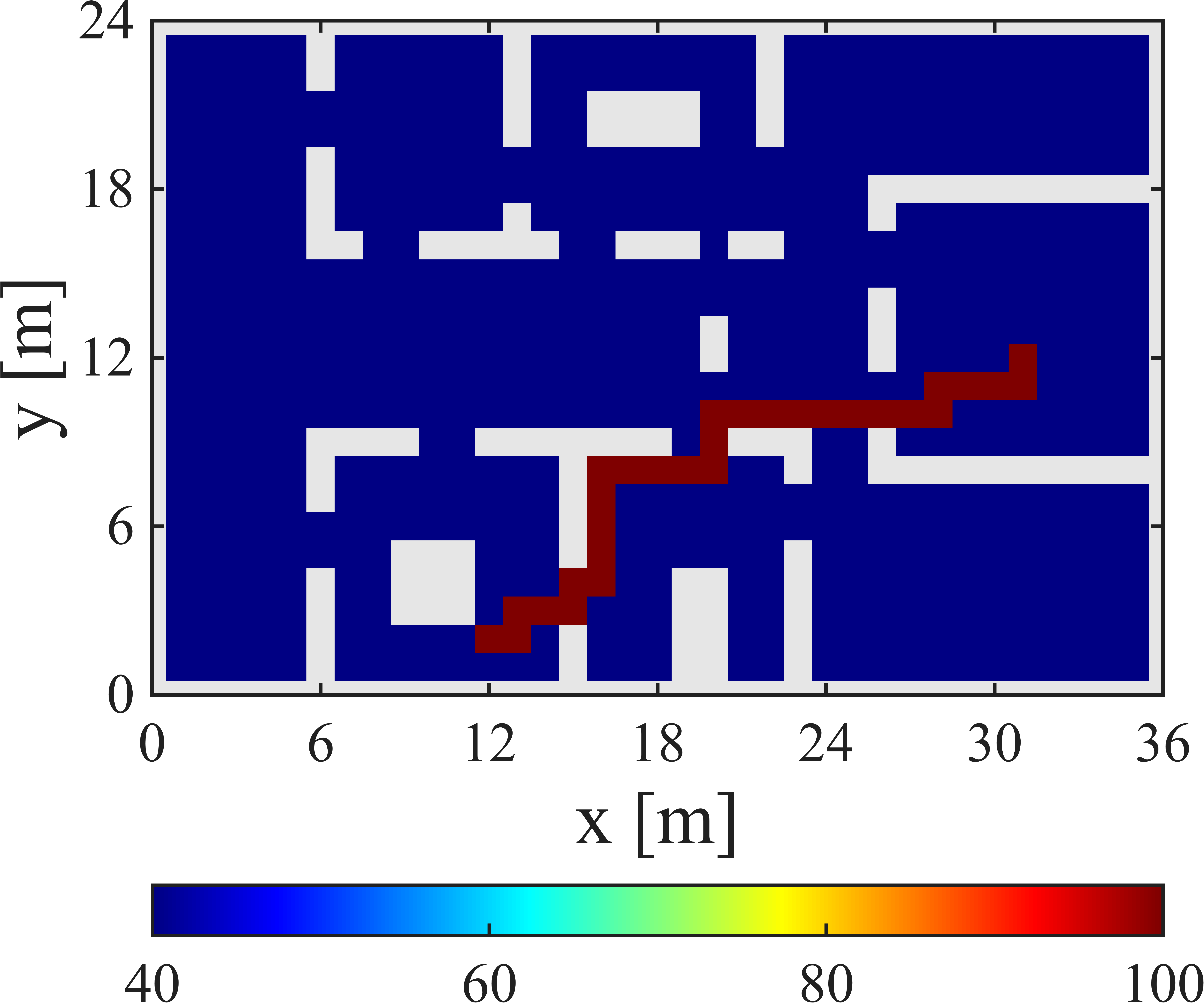}
	}
	
	\caption{Maximum $Q$-values heatmaps $\max_{\mathbf{a}}\,Q_{\text{MF}}(\mathbf{s}_{i,t},\mathbf{a})$ for \ac{PIT} \ac{MF} of each agent at the end of the final training episode. Each column, from left to right, corresponds to a different agent.} 
    \label{fig:max_Q_table}
    \vspace{-15pt}
\end{figure*}

\begin{figure*}[!t]
	\centering
	\subfloat[]{
		\includegraphics[width=0.238\linewidth]{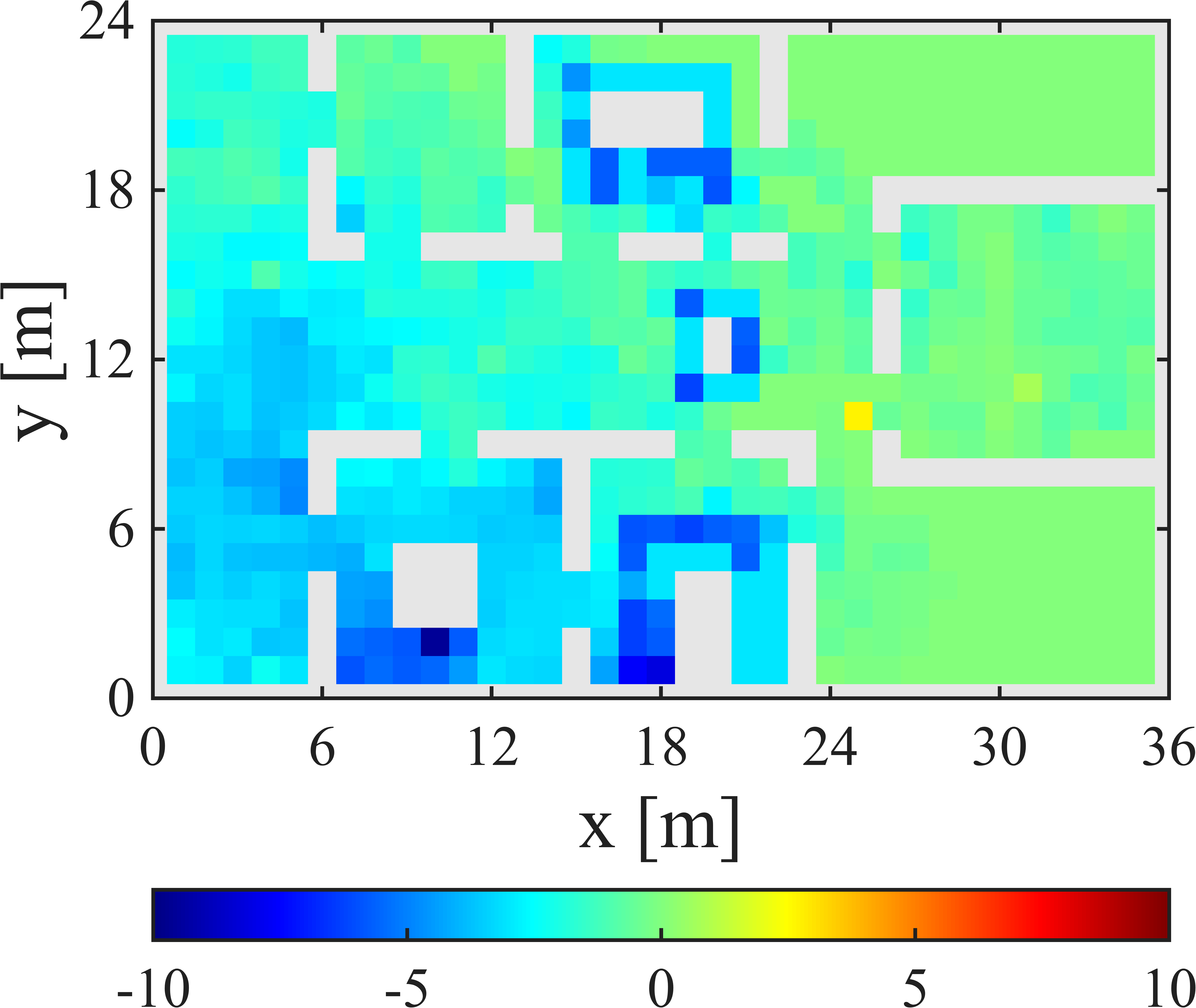}
	}
	\subfloat[]{
		\includegraphics[width=0.238\linewidth]{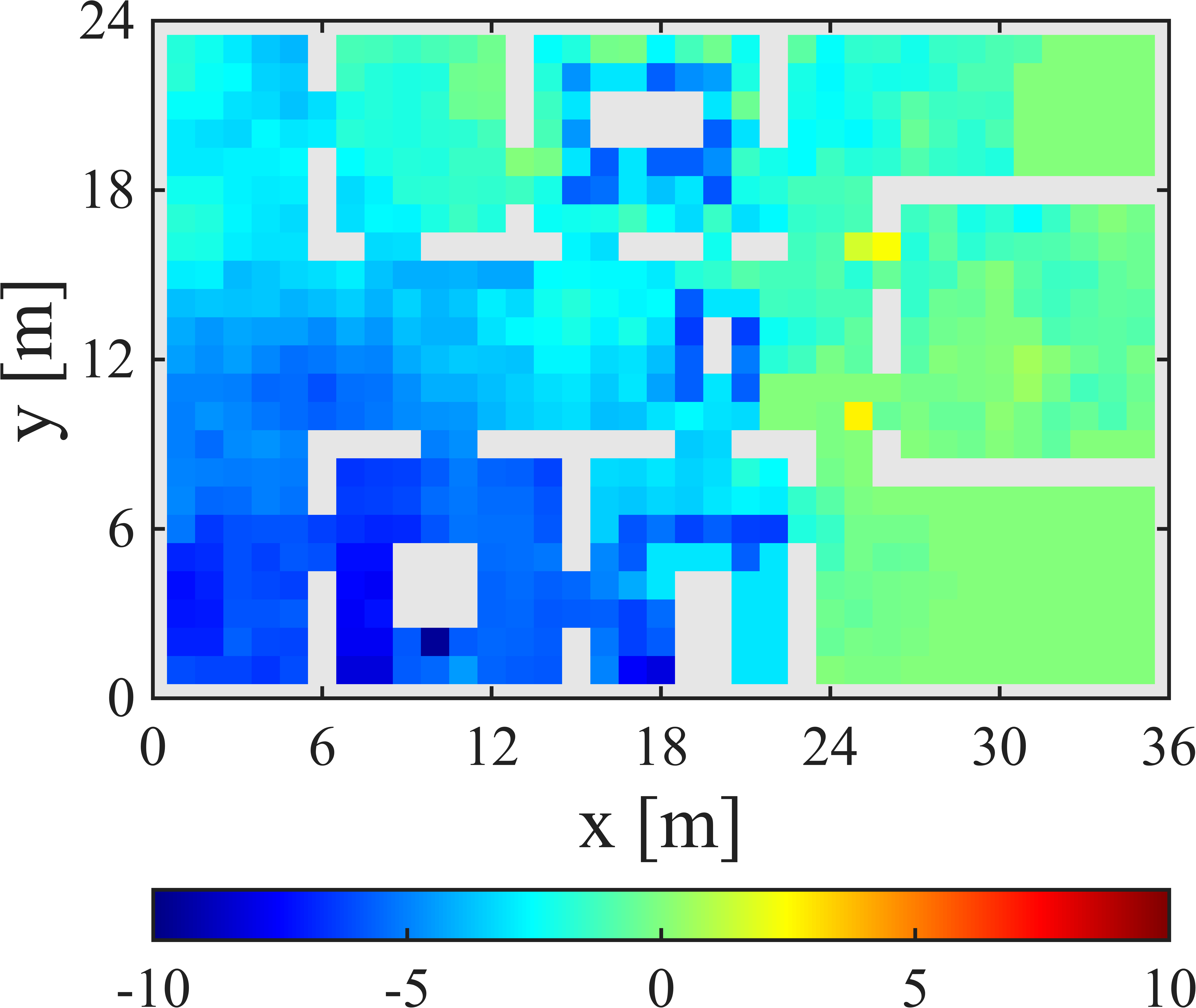}
	}
    \subfloat[]{
        \includegraphics[width=0.238\linewidth]{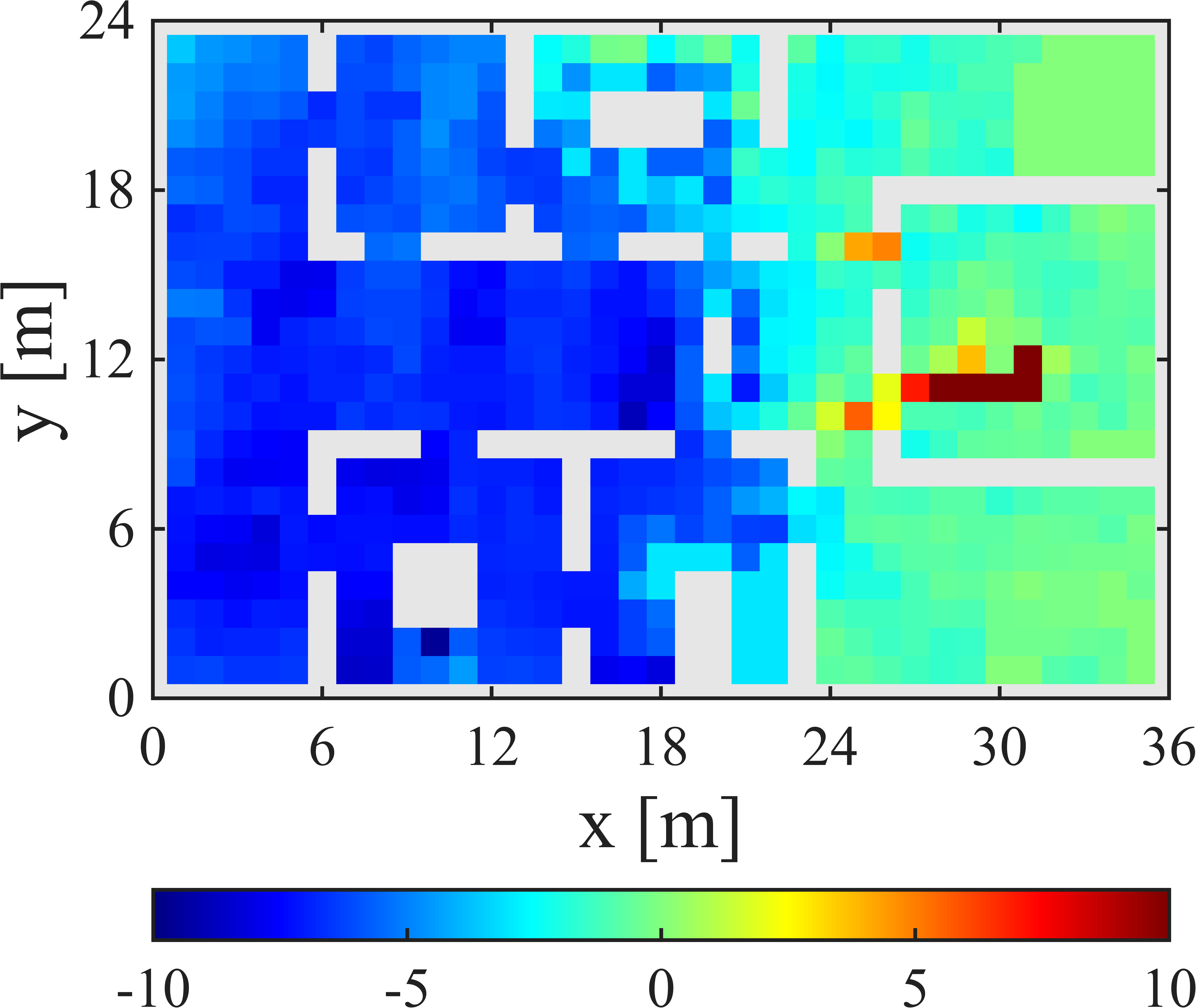}
	}
	\subfloat[]{
		\includegraphics[width=0.238\linewidth]{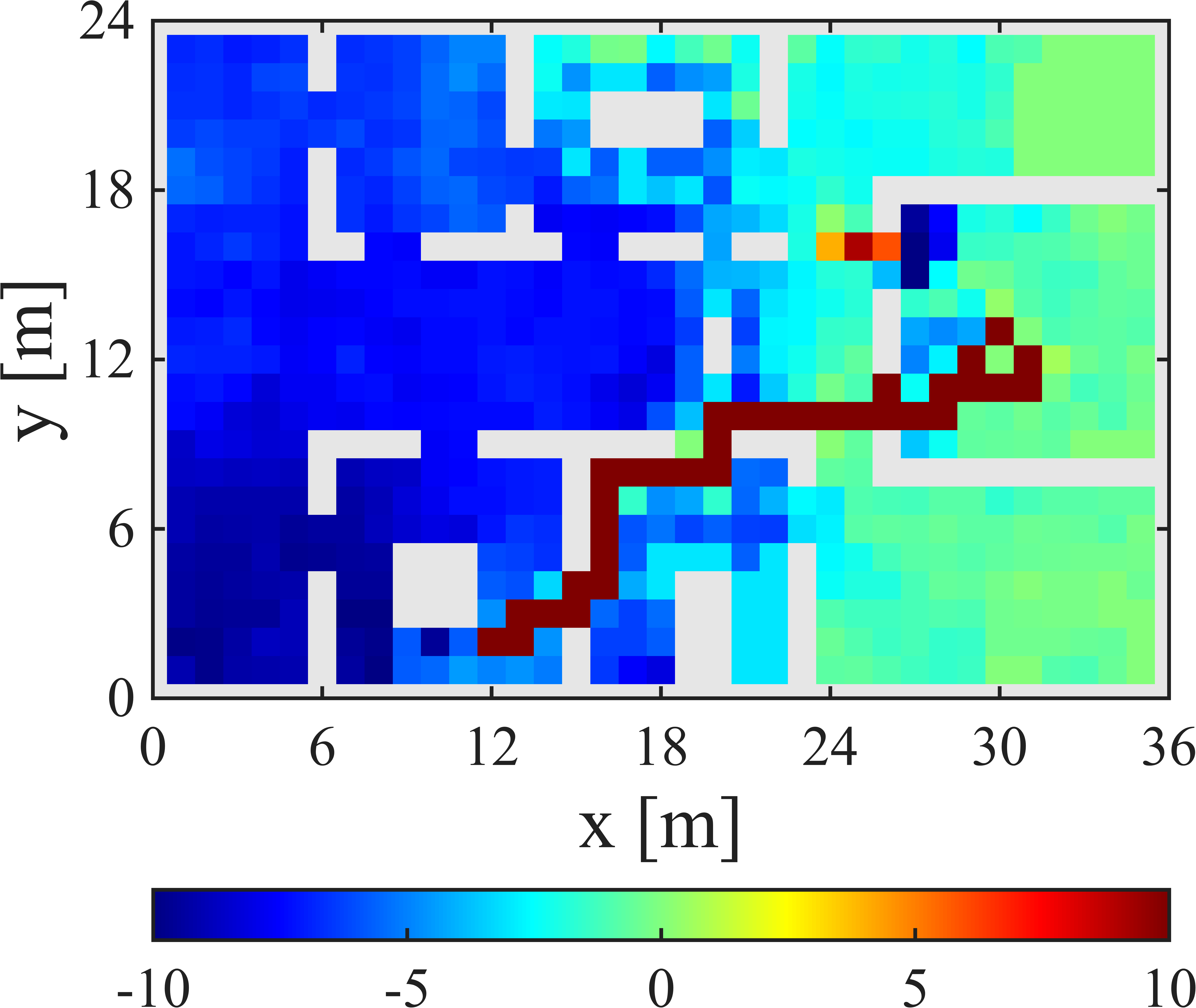}
	}
	
	\caption{Pavlovian value heatmaps of Agent~4 at various stages, namely, Episode $10$, Episode $50$, Episode $100$, Episode $1200$, respectively.} 
    \label{fig:Pavlovian_value}
    \vspace{-20pt}
\end{figure*}

\subsubsection{Instrumental and Pavlovian Values}
The combined value function $\omega(\mathbf{s}_{i,t}, \mathbf{a}_{i,t})$ in \eqref{eq:omega}, computed at the end of training, reveals how the two learning systems cooperate to produce the observed optimal behavior, offering insight into the underlying decision-making process. To examine these contributions, we plot separately the instrumental $Q$-values and the Pavlovian values. 

Fig~\ref{fig:max_Q_table} shows the maximum instrumental action-value for each agent, 
$\max_{\mathbf{a}} Q_{\text{MF}}(\mathbf{s}_{i,t}, \mathbf{a})$. The resulting $Q$-table heatmaps for the directional actions display strong positive values in regions where a given action leads directly toward the high-value area surrounding the target. This structure provides the fine-grained action selection required for the direct paths observed in Fig.~\ref{fig:final_path}. These results confirm that the instrumental learning system successfully estimates the expected cumulative reward associated with reaching the goal from each state, thereby forming a reliable solution that guides the agent toward the extrinsic reward.

Fig~\ref{fig:Pavlovian_value} illustrates the Pavlovian state value $V_4(\mathbf{s})$ (computed via \eqref{eq:pavlovianupdate}) for Agent~4 at several stages of training.  Unlike standard \ac{RL} value functions, which typically encode proximity to the goal, the Pavlovian value evolves into a ``risk–reward potential field" shaped by environmental radio cues. The heatmaps exhibit distinct local maxima at gate locations and local minima in GPS-denied regions. These values act as intrinsic shaping signals; for example, the negative potential in GPS-denied regions reduces the likelihood of selecting actions that enter these regions, effectively pruning the search space for the instrumental learner.
\vspace{-15pt}

\subsection{\ac{PIT} \ac{MF}-\ac{MB} Learning Algorithm: Performance Assessment}

In this subsection, we consider the \ac{PIT} \ac{MF}-\ac{MB} (Hybrid) learning approach, and we compare it with the following baselines:
\begin{itemize}
    \item \emph{Instrumental \ac{MF}}: A standard \ac{MF} \ac{RL} agent using the $Q$-learning update in \eqref{eq:instrumentalupdate} and a softmax policy without Pavlovian modulation. Action selection is driven solely by the instrumental action-value function $Q_{\text{MF}}$.

    \item \emph{\ac{PIT} \ac{MF}}: A \ac{MF} \ac{RL} agent in which action selection is influenced by both the instrumental critic and the Pavlovian value function through the modulation term $g_a(V)$ in \eqref{eq:omega}. This corresponds to the \ac{PIT} \ac{MF} \ac{RL} approach described in Sec.~\ref{sec:methodology}.

    \item \emph{Instrumental \ac{MF}-\ac{MB}}: A Dyna-$Q$ agent that combines the instrumental $Q_{\mathrm{MF}}$ and \ac{MB} $Q_{\mathrm{MB}}$ estimates as in~\eqref{eq:Qhybrid}, but with no Pavlovian contribution.

    \item \emph{\ac{PIT} \ac{MF}-\ac{MB} (Hybrid)}:
    The full proposed algorithm, where the hybrid $Q_{\mathrm{hybrid}}$ in~\eqref{eq:Qhybrid} integrates \ac{MF} and \ac{MB} value estimates, and action selection is further modulated by Pavlovian predictions through $g_a(V)$.
\end{itemize}

These four variants allow us to isolate the individual contributions of
(i) Pavlovian modulation, (ii) \ac{MB} planning, and (iii) their combined effect within the hybrid approach.

\subsubsection{Dynamic Arbitration between \ac{MB} and \ac{MF}}
We first examine the effect of integrating \ac{MB} planning with \ac{MF} learning through the Bayesian arbitration mechanism. This analysis highlights how the balance between MB and MF control evolves during training.

\begin{figure}[t!]
    \centering
    \input{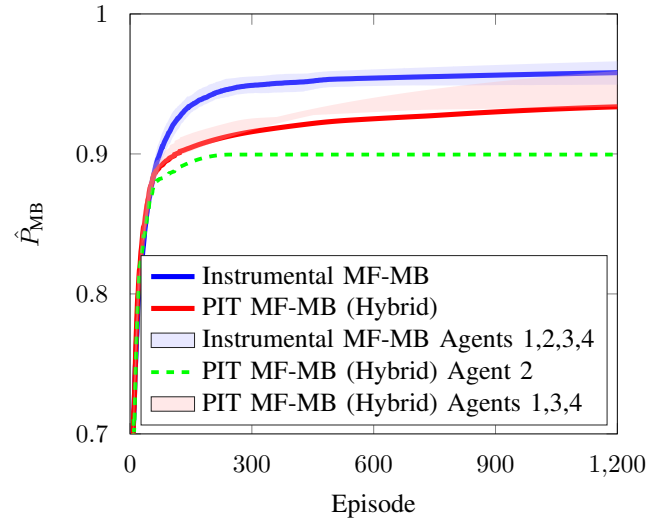}
        
    \caption{ $P_{\text{MB}}$ as a function of episodes. $P_{\text{MB}}$ averaged over the number of agents (denoted as $\hat{P}_{\text{MB}}$) between Instrumental \ac{MF}-\ac{MB} (blue solid line) and the proposed \ac{PIT} \ac{MF}-\ac{MB} (Hybrid) (red solid line); and $P_{\text{MB}}$ between agents with Instrumental \ac{MF}-\ac{MB} and the proposed \ac{PIT} \ac{MF}-\ac{MB} (Hybrid) algorithm. The shaded areas bound the minimum and maximum values across the agents.}
    \label{fig:PMB}
    \vspace{-25pt}
\end{figure}

Fig~\ref{fig:PMB} illustrates the evolution of the arbitration probability $P_{\text{MB}}$, averaged across agents, over the course of training for both the Instrumental \ac{MF}-\ac{MB} and the proposed \ac{PIT} \ac{MF}-\ac{MB} (Hybrid) architectures. This probability reflects the relative reliance placed on the \ac{MB} system and the \ac{MF} system in action selection, as determined by the Bayesian arbitration mechanism described in Section IV-D.

In the early stages of training, both systems exhibit low arbitration probabilities, indicating that neither \ac{MB} nor \ac{MF} control is yet dominant. The reason is the internal models are incomplete and value estimates are still noisy.

As training progresses, a key divergence emerges between two architectures. In the Instrumental \ac{MF}-\ac{MB} baseline, the arbitration probability rises steadily and stabilizes at a relatively high value. This indicates that the \ac{MB} system becomes increasingly favored over the \ac{MF} system. The underlying reason lies in the persistent unreliability of the MF learner: without Pavlovian modulation, the agent frequently explores uncertain or high-risk regions (e.g., GPS-denied areas), leading to high reward prediction errors (\ac{RPE}s). In contrast, the \ac{MB} system, once its transition model becomes sufficiently accurate, yields low state prediction errors (\ac{SPE}s), making it the more reliable controller from the perspective of the arbitrator.

In contrast, the PIT \ac{MF}-\ac{MB} (Hybrid) architecture exhibits a markedly different arbitration profile. Although $P_{\text{MB}}$ also increases during the mid-training phase, it eventually stabilizes at a lower value than that observed for the instrumental baseline. This outcome may initially appear counterintuitive, as one might expect Pavlovian bias to further enhance \ac{MB} reliability. However, it reflects a fundamental property of the proposed hybrid design, that is, Pavlovian conditioning improves the reliability of the MF system itself, thereby reducing the need for \ac{MB} dominance in later stages.

By embedding Pavlovian value signals directly into the action selection policy, the hybrid agent learns to avoid high-risk regions and exploit perceptually informative cues (e.g., gates) early in training. This leads to more consistent and predictable transitions, which in turn stabilizes MF value estimates and reduces \ac{RPE}s. As a result, the MF system becomes a more trustworthy controller, and the arbitrator no longer needs to rely predominantly on the \ac{MB} system. The \ac{MB} planner remains available and accurate, but the arbitration mechanism naturally shifts toward a more balanced configuration when both systems are reliable.

This interpretation is further supported by the agent-specific arbitration profiles. Agents operating in more structured or constrained regions (e.g., Agent~$2$ in Fig~\ref{fig:PMB}) continue to exhibit lower \ac{MB} reliance for longer periods, reflecting its higher reliance for Pavlovian bias guidance. Meanwhile, agents in more open areas (e.g., Agent~$1$) transition more quickly to MB-dominated control. These observations demonstrate that arbitration is a context-sensitive, agent-specific learning process.

\begin{figure}[!t]
	\centering
	\subfloat[]{
		\includegraphics[width=0.9\linewidth]{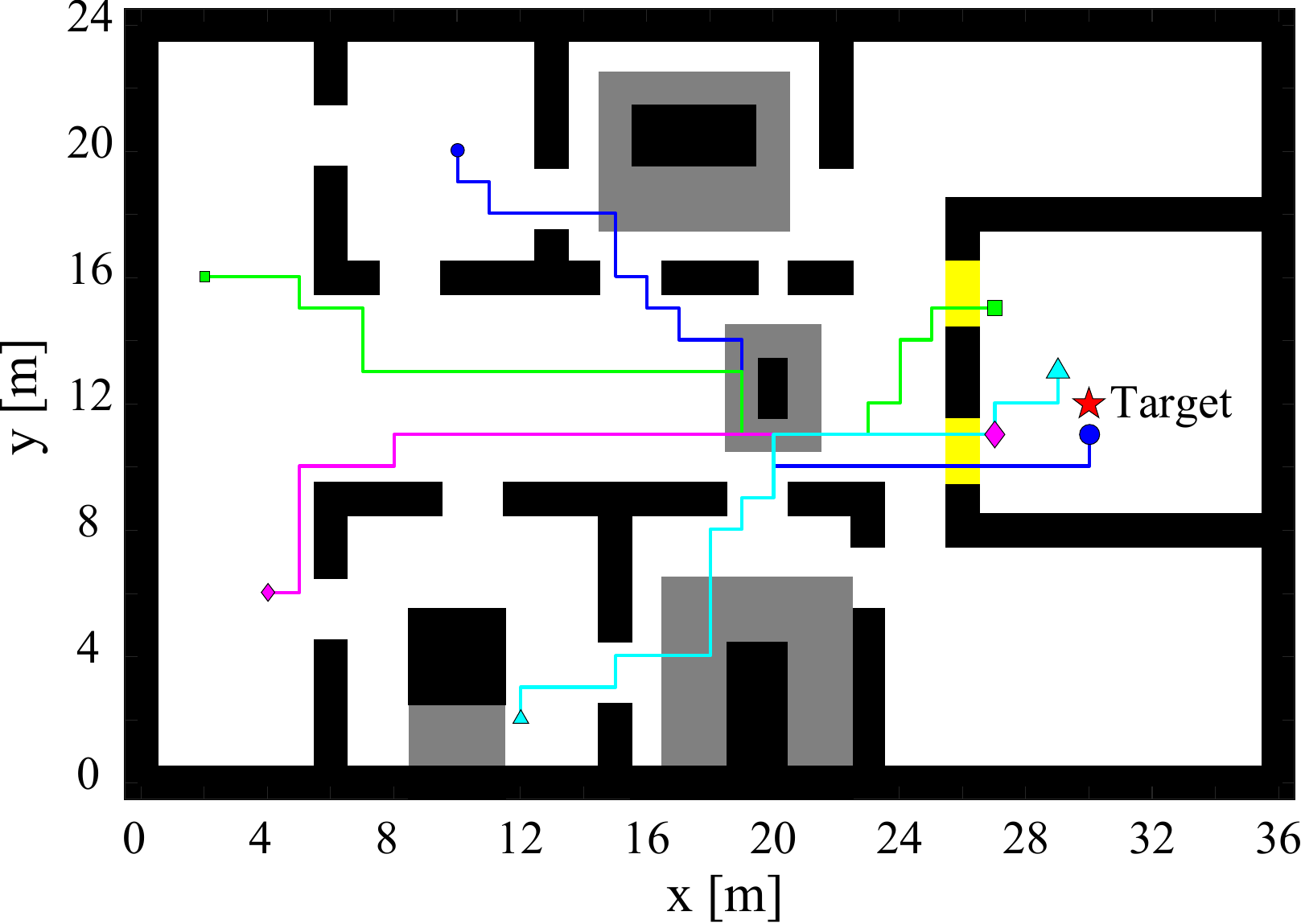}
	}
    
    \subfloat[]{
		\includegraphics[width=0.9\linewidth]{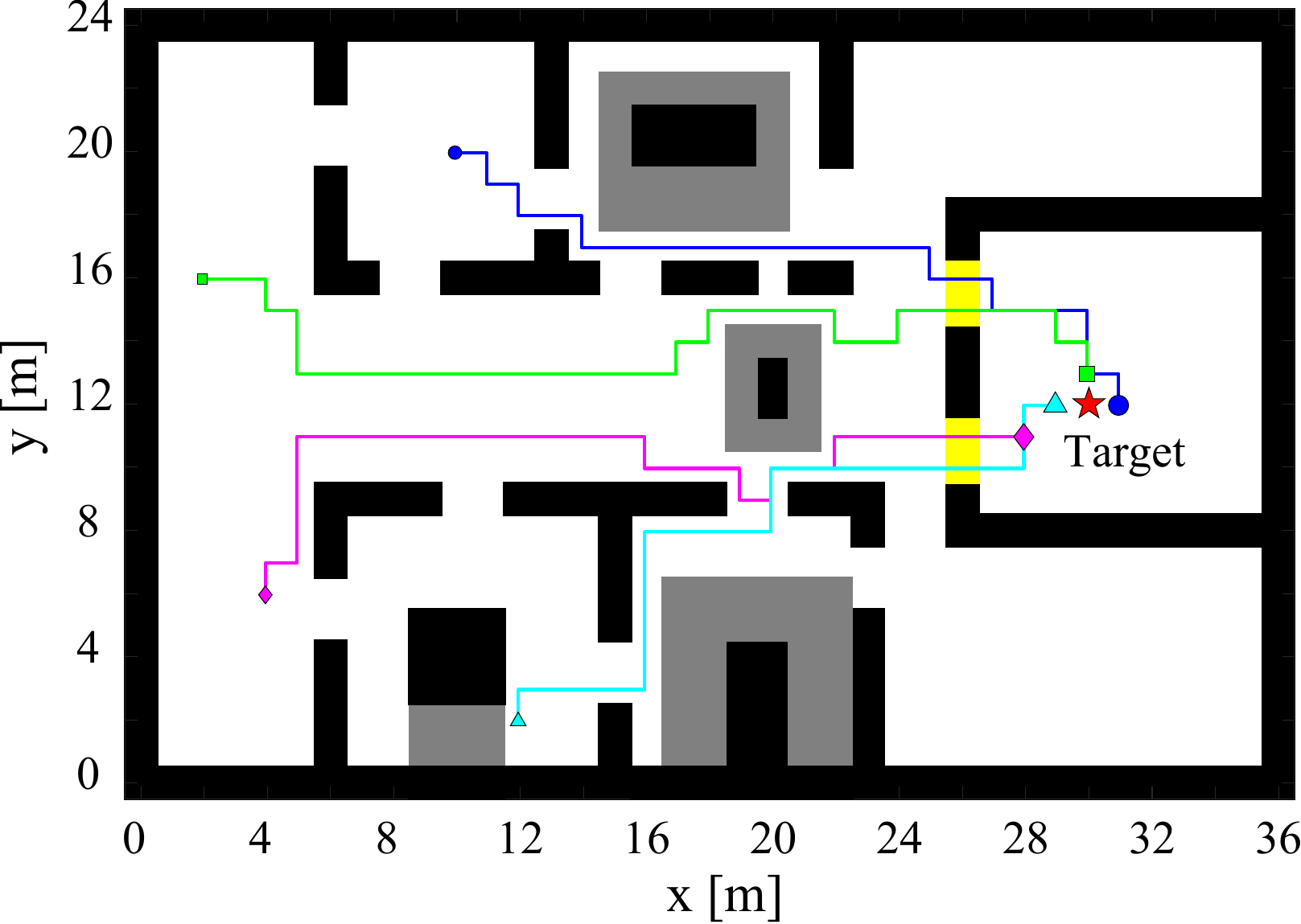}
        \label{fig:path_full}
	}
	\caption{Agent final trajectories with (a) Instrumental \ac{MF}-\ac{MB}, (b) \ac{PIT} \ac{MF}-\ac{MB} (Hybrid) learning approaches.}
    \label{fig:agent_trajectory_modelbased}
    \vspace{-10pt}
\end{figure}
\vspace{-10pt}

\subsubsection{Impact of Pavlovian Modulation to Agents' Navigation}

Fig.~\ref{fig:agent_trajectory_modelbased} compares the final trajectories of the proposed hybrid framework against a standard Instrumental \ac{MF}-\ac{MB} baseline. The baseline agents successfully locate the target but generate paths that graze the GPS-denied region, incurring higher measurement variance. This occurs because the standard $Q$-learning update slowly propagates the negative reward from the GPS-denied region. 

In contrast, the proposed \ac{PIT} \ac{MF}-\ac{MB} approach (Fig. \ref{fig:path_full}) executes a decisive avoidance maneuver well before reaching the hazardous zone. This is a direct manifestation of the \ac{PIT} phenomenon. GPS-denied regions act as \ac{CS} predicting punishment, leading to higher \ac{PEB} values.  Through the action-dependent modulation, any action leading toward these states is penalized in the softmax logits, creating a repulsive gradient. This cue-driven bias operates immediately upon perceiving the predictive region, enabling rapid, reflex-like avoidance without requiring expensive trial-and-error learning of the long-term consequences.

\begin{figure}[t!]
    \centering
    \input{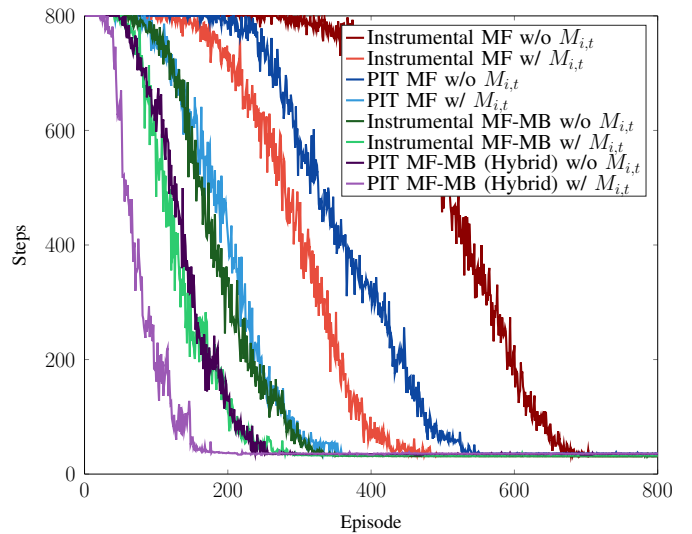}
    \caption{Learning rates expressed as steps per episodes and for various combinations of learning types. ``w/" and ``w/o" refer to the presence, or not, of the motivational signal.}
    \label{fig:step_comparison}
    \vspace{-15pt}
\end{figure}

\subsubsection{Comparative Evaluation}
Finally, having characterized the individual roles of Pavlovian modulation and \ac{MB} arbitration, we compare the performance of all algorithmic variants.
This comparison isolates the contribution of each subsystem, Pavlovian, \ac{MF}, and \ac{MB}, and quantifies their combined effect within the proposed hybrid architecture.

Fig. \ref{fig:step_comparison} shows the number of steps needed to achieve the \ac{PEB} goal as a function of the number of episodes, averaged over $40$ Monte Carlo simulations. Results are shown for four algorithmic configurations, each presented both with and without Pavlovian modulation. We can observe that the proposed \ac{PIT} \ac{MF}-\ac{MB} (hybrid) approach (e.g., the black solid line) reaches a competent policy significantly earlier than Instrumental \ac{MF} (e.g., the red solid line). This confirms the hypothesis that a cue-driven learning process accelerates exploration toward low-risk, information-rich regions. We can also observe that \ac{MB} approaches produce a more deterministic final policy faster than Instrumental \ac{MF} and \ac{PIT} \ac{MF} approaches. The hybrid \ac{PIT}  \ac{MF}-\ac{MB} algorithm achieves the fastest stabilization of around $200$ episodes. 

The results above demonstrate that integrating Pavlovian, instrumental, and \ac{MB} components yields capabilities that none of the individual subsystems can achieve on its own. The Pavlovian modulation provides immediate, cue-driven shaping that accelerates exploration toward perceptually informative structures, while simultaneously discouraging transitions into geometrically hazardous areas such as GPS-denied regions. This shaping substantially reduces the burden on the instrumental learner and leads to faster discovery of high-value areas, such as \ac{LOS} regions. The instrumental component contributes fine-grained action discrimination and long-term reward estimation, supplying the precise directional gradients required for efficient multi-agent localization. The \ac{MB} planner further enhances stability by propagating consistent predictions, smoothing local inconsistencies in instrumental values, and reducing dependence on noisy or incomplete real-world experience. The arbitration mechanism coordinates these elements dynamically, relying on the planner only when its internal model is reliable, and allowing the \ac{MF} system to dominate when rapid, stable exploitation is sufficient. Together, these findings indicate that the cognitive-inspired hybrid \ac{PIT} \ac{MB} and \ac{MF} architecture provides a principled approach to enhance autonomous decision-making in complex environments.

Moreover, across all architectures, the inclusion of the motivational signal consistently accelerates convergence, as evidenced by the lower number of steps required to reach the PEB goal. This effect is particularly pronounced in the Instrumental \ac{MF} and \ac{PIT} \ac{MF} configurations The performance gap underscores the role of the motivational signal as an internal drive mechanism that dynamically adjusts behavior based on resource depletion and mission urgency. By introducing a state-dependent cost term, the motivational signal penalizes inefficient actions and biases exploration toward goal-relevant regions, thereby reducing time spent in low-utility areas. Notably, even in the \ac{PIT} \ac{MF}-\ac{MB} (Hybrid) architecture, which already benefits from Pavlovian biasing and \ac{MB} planning, and the addition of the motivational signal yields further improvement, suggesting that internal drive modulation provides a complementary and non-redundant contribution to learning. These results corroborate the biological inspiration underlying the proposed framework: just as hunger or fatigue shapes decision-making in humans, the motivational signal enables autonomous agents to adapt their exploration-exploitation trade-off under internal constraints, leading to faster and more resource-efficient navigation.
\vspace{-10 pt}

\section{Conclusions}

In this work, we introduced a human-inspired hybrid \ac{RL} framework that unifies Pavlovian conditioning, instrumental \ac{MF} learning, and \ac{MB} planning to enhance autonomous agents' navigation in uncertain environments. We introduced the concept of \emph{contextual radio cues}, defined as specific georeferenced states of the environment acting as conditioned stimuli that predict task-relevant outcomes (i.e., increases or decreases in the PEB) based on their inherent electromagnetic characteristics. Then, leveraging the principle of \ac{PIT}, the framework uses these contextual radio cues to shape action selection by biasing the instrumental-only policy using an action-dependent value function. This \ac{PIT}-based policy enables agents to exploit contextual information to approach task-informative regions and proactively avoid hazardous areas.

A Bayesian arbitration mechanism further stabilizes behavior by dynamically 
balancing habitual control (\ac{MF} learning) and deliberative planning 
(\ac{MB} control) according to reliability scores. Each agent performs 
this arbitration locally based on its own experience and internal signals. 
Although agents learn independently and do not explicitly share learned 
value functions or policies, they interact through the shared environment 
during the localization task. Simulation results for a multi-agent 
localization task show that this integrated design accelerates convergence 
and improves safety compared to each subsystem operating in isolation.

Future works will extend the architecture to high-dimensional continuous state-action spaces through deep \ac{RL}, strengthen multi-agent cooperation through communication of Pavlovian values and internal model updates, and validate the approach on physical platforms to evaluate robustness under real-world sensing and environmental variability.
\vspace{-10pt}

\label{sec:conclusions}

\bibliographystyle{IEEEtran}

\end{document}